\documentclass[aps,epsfig,rotate,showpacs]{revtex4}
\usepackage{epsfig}
\usepackage{graphicx}
\usepackage{amsmath}

\newcommand{\beq}{\begin{eqnarray}}
\newcommand{\eeq}{\end{eqnarray}}

\def\qi{{\bf \overline{\gamma}}}
\def\qj{{\bf \gamma}}
\begin{document}                
\title{
Chaos and Quantum-Classical Correspondence via Phase Space Distribution Functions}
\author{Jiangbin Gong \cite{bylineone} and Paul Brumer}
\affiliation{Chemical Physics Theory Group,Department of Chemistry,
University of Toronto, Toronto, Canada  M5S 3H6}
\date{\today}

\begin{abstract}
Quantum-classical correspondence in  conservative chaotic Hamiltonian
systems is examined using a uniform structure measure for quantal and
classical phase space distribution functions. The similarities and
differences between quantum and classical  time-evolving distribution
functions are exposed by both analytical  and  numerical means. The
quantum-classical correspondence of low-order statistical moments is also
studied.  The results shed considerable light on quantum-classical
correspondence.
\end{abstract}

\maketitle

\section{Introduction}
\label{2-1}

The quantum dynamics of systems that are classically chaotic has been a
subject of considerable interest for nearly three decades. Chaos, usually
defined as the exponential sensitivity of phase space {\em trajectories}
to slight changes of the initial conditions, resists direct translation to
the natural Hilbert space setting of quantum mechanics since there do not
exist well-defined trajectories within the conventional interpretation of
quantum mechanics \cite{exponential}.   Hence, many studies on quantum chaos in the
literature have been dedicated to the relatively modest study of quantum
manifestations of classical chaos \cite{book1,book2}. Actual
quantum-classical correspondence (QCC) in classical chaotic systems is
still one of the outstanding  issues in quantum theory.

Considerations of quantum-classical correspondence that rely upon the standard
formulations of quantum and classical mechanics are at a great disadvantage.
Specifically, classical mechanics does not admit wavefunctions and quantum
mechanics does not admit trajectories. Hence one is faced with trying to
connect two theories which do not even have the same basic rudimentary elements.

A far more enlightening approach is to utilize the Hilbert space
formulation of both quantum \cite{vonNeumann}  and classical \cite{Koopman}
mechanics where the central element is the density operator in both
formulations. By choosing to deal with the phase space representation of
quantum mechanics one can then compare the classical and quantum dynamics
of distributions in phase space. Indeed, all the associated machinery of
commutation rules, eigenvalues, eigenstates, etc. can be used to formally
analyze quantum-classical correspondence of both integral and chaotic
systems \cite{wilkie}.

Recognizing that phase space distributions allow for a common view of
classical and quantum mechanics (see also
\cite{fox,ballentine94,ballentine01,chris}) suggests that we consider the
nature of chaos as it manifests itself in phase space distributions, as
distinct from classical phase space trajectories. 
To this end we \cite{arjendu1}, and
others \cite{gu,partovi}, have developed a criteria for chaos in terms of the
structure of phase space distribution functions.

In this paper, the distribution function approach to QCC in classically
chaotic systems is carefully examined and substantially extended. The
extension is from one-dimensional chaotic systems in previous work
\cite{gu,arjendu1} to two-dimensional conservative systems, from the
special case of uniform dynamical instabilities to the general case of
nonuniform stretching and contraction rates fluctuating with time and
phase space location. That is, armed with a measure of the structure of
phase space distribution functions, it becomes possible to quantitatively
investigate the similarities and differences between quantal and classical
distribution functions in a broad class of smooth, conservative and
strongly chaotic Hamiltonian systems. In particular, this paper displays
(1) the rich transient behavior of the dynamics of distribution functions
shared by quantum and classical dynamics before a QCC break time $t_{b}$,
a quantity which scales with the logarithm of $\hbar$, (2) a simple but
enlightening description of the break regime of QCC in the time
development of phase space structure, and (3) interesting QCC in low order
statistical moments during a complex relaxation process with a time scale
much larger than $t_{b}$.

We note in passing that  the distribution function strategy is also of
considerable interest to the fundamental understanding of decoherence in
quantum vs. classical mechanics. For example,  the structure of the
classical or quantum distribution functions determines properties of the
system when coupled to an environment \cite{arjendu2,arjendu3}. Hence, QCC
from a distribution function viewpoint is useful to the understanding
of short-time quantum decoherence rates versus rates of classical entropy
production \cite{gongprl}. Furthermore, QCC between quantal and classical
distribution functions is closely related to the issue of decoherence in
the presence of an environment that has a chaotic classical limit
\cite{zurekprl02}.

The model calculations in this paper are based on a strongly chaotic
system: the quartic oscillator model \cite{Eckhardt}. The Hamiltonian is, in
dimensionless scaled variables \cite{gong}, given by
\beq
H(q_{1}, q_{2}, p_{1}, p_{2})=\frac{p_{1}^{2}}{2}+\frac{p_{2}^{2}}{2}+\frac{\alpha}{2}
q_{1}^{2}q_{2}^{2}+\frac{\beta}{4}(q_{1}^{4}+q_{2}^{4}).
\label{hamiltonian2}
\eeq
When $\alpha=\beta$ or $3\beta$, this system is completely integrable.  For
very large values of $\alpha/\beta$ the system is strongly chaotic,  with
the characteristic Lyapunov exponent being an order of
magnitude larger than that of other
conservative chaotic systems, e.g.,  the Henon-Heiles system.

This paper is organized as follows. Section \ref{s2-2} briefly reviews the
distribution function approach to QCC in chaotic systems. Special emphasis
is put on a structure measure of classical and quantal distribution
functions, and on our definitions of classical and quantal finite-time
Lyapunov exponents. The paper then proceeds to present results in order of
increasing time scales. In Sec. \ref{s2-3}, QCC is studied for initially
positive-definite Wigner functions at early times. In Sec. \ref{s2-4}, a
simple analytical understanding of the break time regime of QCC is
provided, followed by supporting numerical results. We then consider, in
Sec. \ref{s2-5}, QCC in some low-order statistical moments for much larger
time scales.  A brief summary is given in Sec. \ref{s2-6}.

\section{Lyapunov Exponents  in Distribution Function Dynamics}
\label{s2-2}
\subsection{Classical distribution function dynamics}

Classical chaos is usually defined as the extreme sensitivity of
trajectories to slight changes in initial conditions.  Quantitatively, it
is described by a positive maximum Lyapunov exponent or by the Kolomogorov
entropy. Consider a conservative Hamiltonian system with two degrees of
freedom with dimensionless canonical variables $q_{1}, q_{2}, p_{1},
p_{2}$. A phase space point is characterized by a $4$-dimensional column
vector $\qj\equiv (q_{1}, q_{2}, p_{1}, p_{2})^{\dagger}$. For brevity we
introduce the antisymmetric matrix ${\bf J}= \left(
 \begin{array}{ll}
  {\bf  0} & {\bf  1}\\ -{\bf 1} & {\bf 0}
      \end{array}
       \right)$
  where ${\bf 0}$ and ${\bf 1}$ are $2\times 2$ zero and unit matrices, respectively.
The canonical equations of motion (i.e., Hamilton's equations)
then read as
$\dot{\qj}={\bf J}\partial H/\partial\qj$.
The sensitivity of classical trajectories to initial conditions
is described by the stability matrix $
{\bf M}_{ij}[\qj(0),t]\equiv
\partial \qj_{i}(t)/\partial \qj_{j}(0)$, and the
maximal Lyapunov exponent $\lambda$ is given by
\begin{eqnarray}
\lim_{t\rightarrow\infty}\frac{1}{t}
\ln\left(\left|{\bf M}[\qj(0), t] {\bf \eta}\right|\right)=\lambda[{\bf \gamma}(0)],
\label{definition}
\end{eqnarray}
where ${\bf \eta}$ is a vector in the tangent space.
The stability matrix ${\bf M}$ is  symplectic and  its time evolution is
governed by the differential equation
\beq
\dot{{\bf M}}={\bf J}\frac{\partial^{2}H}{\partial {\bf \gamma}^{2}}{\bf M},
\label{Mt}
\eeq
with
\beq
{\bf M}(0)=1.
\label{m0}
\eeq

A simple application of Liouville's theorem leads to an alternative
expression for the maximal Lyapunov exponent in terms of classical
distribution function dynamics \cite{gu,arjendu1}. Let $\rho_{t}$ denote a
well behaved classical probability distribution and ${\bf \xi}$ denote an
arbitrary infinitesimal vector in phase space. Then, from the
incompressibility of the Liouville density function, one has
\beq
\rho_{t}[{\bf \gamma}(t)]+{\bf \nabla} \rho_{t}
[{\bf \gamma}(t)]{\bf M\xi}=
\rho_{0}[{\bf \gamma}(0)]+{\bf \nabla } \rho_{0}[{\bf \gamma}(0)]\xi.
\eeq
Further, using
${\bf J}^{2}=-1$ and $ {\bf JM}^{\dagger}{\bf J}=-{\bf M}^{-1}$,
one obtains
\begin{eqnarray}
{\bf MJ}{\bf \nabla}
\rho_{0}[{\bf \gamma}(0)]=
{\bf J}{\bf \nabla}\rho_{t}[{\bf \gamma}(t)].
\label{mpm}
\end{eqnarray}
Substituting Eq. (\ref{mpm}) into Eq. (\ref{definition}) yields
\beq
\lambda[\qj(0), t]& = &\lim_{t\rightarrow \infty}\frac{1}{t}\ln\left|{\bf J}{\bf \nabla}
\rho_{t}[{\bf
\gamma}(t)]\right| \nonumber \\
& = & \lim_{t\rightarrow \infty}\frac{1}{t}\ln\left|{\bf \nabla}
\rho_{t}[{\bf
\gamma}(t)]\right|.
\label{maxlambda}
\eeq
Note that Eq. (\ref{maxlambda}) defines the Lyapunov exponent in terms of
phase space distribution properties. {\em Specifically, the faster the
structure of} $\rho_{t}$ increases, {\em the larger the} $\lambda$.

In accord with Ref. \cite{arjendu1}, we define a particular ensemble
average of the gradients of distribution functions as $\chi_{2c}$,  i.e.,
\beq
\chi_{2c}(t)\equiv \left[\frac{\int |{\bf \nabla}\rho_{t}({\bf \gamma})|^{2}d{\bf \gamma}}{
\int \rho_{t}^{2}({\bf \gamma})d{\bf \gamma}}\right]^{1/2}=\left[
-\frac{\int \rho_{t}({\bf \gamma}) {\bf \nabla}^{2}\rho_{t}({\bf \gamma})d{\bf \gamma}}{
\int \rho_{t}^{2}({\bf \gamma})d{\bf \gamma}}\right]^{1/2}.
\label{definechi}
\eeq
Using Eqs. (\ref{mpm}) and (\ref{definechi}), we have
\beq
\chi_{2c}(t)=\left[\frac{1}{\int\rho_{0}^{2}({\bf \gamma)}d{\bf
\gamma}}\int \left|{
\bf JM}(\qj,
t){\bf J}
{\bf \nabla}
\rho_{0}({\bf \gamma})\right|^{2}d{\bf \gamma}\right]^{1/2}.
\label{chi2t}
\eeq

The physical significance of $\chi_{2c}(t)$ becomes apparent when one
considers the Fourier transform of the distribution function. That is,
suppose $\rho_{t}({\bf \gamma})=[1/(2\pi)^{4}]\int d{\bf k} \exp(2\pi i
{\bf k}\cdot {\bf \gamma})\rho_{t }({\bf k})$,  where $\rho_{t}({\bf k})$
denotes the Fourier component evaluated at the $4$-dimensional wave vector
 ${\bf k}$. Then
\beq
 \chi_{2c}^{2}(t)=\frac{\int d{\bf k} {\bf k}^{2}|\rho_{t}({\bf k})|^{2}}{\int d{\bf
 k}|\rho_{t}({\bf
  k})|^{2}},
\eeq
showing that $\chi_{2c}$ is the root-mean-square radius of the Fourier
transform of the distribution  function, and thus serves as a measure of
classical phase space structure. That is, the larger the $\chi_{2c}$, the
more structured is $\rho_t(\gamma)$.

For completely integrable dynamics there exists a special set of
generalized coordinates: action variables $I_{1}, I_{2}$ and angle
variables $\theta_{1}, \theta_{2}$. In this representation, the
Hamiltonian depends only on the action variables that are constants of
motion. For such cases, Eq. (\ref{Mt}) has the simple solution
\begin{eqnarray}
{\bf M}=1+{\bf J}\frac{\partial^{2}H}{\partial {\gamma}^{2}}t.
\end{eqnarray}
Substituting this explicit time dependence of the stability matrix
into Eq. (\ref{chi2t}) gives
the following long-time behavior of $\chi_{2c}$:
\begin{eqnarray}
\lim_{t\rightarrow \infty}\frac{\chi_{2c}(t)}{t}=\left[\frac{1}{\int \rho_{0}^{2}({\bf \gamma}
)d{\gamma
}}\int \left|
\frac{\partial^{2}H}{\partial {\gamma}^{2}}{\bf J}{\bf \nabla} \rho_{0}({\bf \gamma})
\right|^{2}d{\bf \gamma}\right]^{1/2}.
\end{eqnarray}
Hence, for regular dynamics the structure of distribution functions, as
measured by  $\chi_{2c}$, asymptotically shows a linear time dependence in
the action-angle representation. However, $\chi_{2c}(t)$ may show a
polynomial time dependence in other canonical representations.

By contrast, for chaotic dynamics it was shown that \cite{arjendu1}
\begin{eqnarray}
\lim_{t\rightarrow \infty}\frac{1}{t}
\ln\chi_{2c}(t)
=\lambda_{2},
\label{limitdefine}
\end{eqnarray}
where $\lambda_{2}$ is the so-called second order generalized maximal
Lyapunov exponent. That is, in the chaotic case the root-mean-square
Fourier radius of distribution functions increases, asymptotically,  at an
exponential rate of $\lambda_{2}$.   Since a given resolution limit
$\delta$ corresponds to the inability to account for Fourier modes larger
than $1/\delta$, chaos can be understood as a kind of exponential loss of
accuracy, or of information, encoded in the Fourier basis expansion of the
initial distribution function.

By definition, the (generalized) Lyapunov exponent $\lambda_{2}$ is an
asymptotic property, relevant as time goes to infinity. Realistically, 
however, it is finite-time
properties of classical chaotic dynamics that are of real interest to the
study of QCC.  
To this end it is useful to introduce a finite-time Lyapunov
exponent. Based upon Eq. (\ref{limitdefine}), we define the finite-time
Lyapunov exponent in terms of the average exponential increase rate of
$\chi_{2c}(t)$ over time $t$:
 \beq
\lambda_{2c}(t)\equiv\frac{1}{t}\ln\left[\frac{\chi_{2c}(t)}{\chi_{2c}(0)}\right],
\label{fdefine}
\eeq
with
\beq
\lim_{t\rightarrow +\infty}\lambda_{2c}(t)=\lambda_{2}.
\label{lambdainfty}
\eeq
We examine this quantity over various time scales later below.

\subsection{Quantal analog of classical finite-time  Lyapunov exponents}

From the ensemble point of view,  QCC is best understood by comparing the
classical Liouville equation with the quantum von Neumann equation in a
phase space representation, e.g., the  Wigner-Weyl representation.
Specifically, given Eq. (\ref{chi2t}), which provides a quantitative
diagnostic for characterizing classical chaos using classical distribution
functions, it becomes straightforward to define the quantum analog of
classical Lyapunov exponents using quantal distribution functions. By
analogy with Eq. (\ref{definechi}) we define
the measure $\chi_{2q}$ for quantal phase space structure as
\beq
\chi_{2q}\equiv \left[\frac{\int |{\bf \nabla}\rho^{W}({\bf \gamma})|^{2}d{\bf \gamma}}{
  \int (\rho^{W})^{2}({\bf \gamma})d{\bf \gamma}}\right]^{1/2},
\label{definechiq}
\eeq
where $\rho^{W}({\bf \gamma})$ is the Wigner function of a quantum state.
Accordingly, in the Fourier space of quantal distribution function,
\beq
\chi_{2q}^{2}=\frac{\int d{\bf k} {\bf k}^{2}|\rho^{W}_{t}({\bf k})|^{2}}{\int d{\bf
 k}|\rho^{W}_{t}({\bf
   k})|^{2}},
   \label{qchi2k}
\eeq
where $\rho^{W}_{t}({\bf
\gamma})=[1/(2\pi)^{4}]\int d{\bf k} \exp(2\pi i {\bf k}\cdot
{\bf \gamma})\rho^{W}_{t
}({\bf k})$.
Hence, $\chi_{2q}$ is the root mean-square Fourier
radius of the Wigner function.
Further, finite-time Lyapunov exponents  $\lambda_{2q}(t)$ for
quantum distribution
function dynamics can
be defined by direct analogy to $\lambda_{2c}(t)$, i.e.,
\beq
\lambda_{2q}(t)\equiv\frac{1}{t}\ln\left[\frac{\chi_{2q}(t)}{\chi_{2q}(0)}\right].
\eeq

Interestingly,   $\chi_{2q}$ has an equivalent expression that is easier
to handle. Suppose $\hat{\rho}$ is the density-matrix operator associated
with the Wigner function $\rho^{W}$, and $\hat{\qj}_{i}$ is the operator
associated with the classical canonical variable $\qj_{i}$ (e.g. $q_i,
p_i$). A simple calculation \cite{arjendu1,gu} then shows that
\beq
\chi_{2q}^{2}=
2\sum_{i}\frac{Tr(\hat{\rho}^{2}\hat{\gamma}_
{i}^{2}
-\hat{\rho}\hat{\gamma}_{i}\hat{\rho}\hat{\gamma}_{i})}{\hbar^{2}Tr(\hat{\rho}^{2})}.
\label{bound}
\eeq
Of particular interest is the pure state case, in which
$\hat{\rho}^{2}=\hat{\rho}$
and as a result,
\beq
\chi_{2q}^{2}=\frac{2}{\hbar^{2}}\sum_{i}(\langle \hat{\gamma}_{i}^{2}\rangle
-\langle
\hat{\gamma}_{i}\rangle^{2}),
\label{chi2q}
\eeq
where $\langle\cdot\rangle$ represents ensemble expectation values. As
shown later below, the 
analogous classical expression is far more complex.

Consider then  the quantum counterpart of Eq. (\ref{lambdainfty}).
Since  Eq. (\ref{bound}) indicates that
$\chi_{2q}^{2}\le 2\sum_{i}Tr(\hat{\rho}^{2}\hat{\qj_{i}}^{2})/
\hbar^{2}Tr\hat{\rho}^{2}$, we have that
$\chi_{2q}$ has an upper bound for any bounded Hamiltonian system. As such,
for fixed $\hbar$ and  bounded systems,
\beq
\lim_{t\rightarrow+\infty}\lambda_{2q}(t)=0.
\label{lambdaqinfty}
\eeq
This reproduces the widely accepted result that bounded quantum systems
cannot exhibit chaos in the strict sense.

However, as noted above, what is of interest to the study of QCC are the
transient properties of $\lambda_{2q}(t)$ versus  $\lambda_{2c}(t)$.
Consider, for example,  a two-degree-of-freedom system $H=H(q_{1}, q_{2},
p_{1}, p_{2})$. The quantum von Neumann equation in terms of the Wigner
function $\rho^{W}$ is given by
\beq
\frac{\partial \rho^{W}}{\partial t}=\{H, \rho^{W}\}+\sum_{(l_{1}+l_{2})>1,\ odd}
\frac{(\hbar/2i)^{(l_{1}+l_{2}-1)}}{l_{1}!l_{2}!}
\frac{\partial^{(l_{1}+l_{2})}V(q_{1},q_{2})}{\partial q_{1}^{l_{1}}\partial q_{2}^{l_{2}}}
\frac{\partial^{(l_{1}+l_{2})} \rho^{W}}{\partial p_{1}^{l_{1}}\partial
p_{2}^{l_{2}}},
\label{qliouville}
\eeq
where the first term on the right hand side is the classical Poisson bracket, and the second
term represents the sum over an infinite series of ``quantum corrections''.  Consider the
short time limit of $\lambda_{2q}(t)$.
Using Eqs. (\ref{definechiq}) and (\ref{qliouville})
we have
\beq
\lambda_{2q}(0) &=&
\frac{1}{\int|{\bf \nabla}\rho_{0}^{W}|^{2} d\qj}\left[\int
({\bf \nabla}
\rho_{0}^{W})^{T}(\frac{\partial^{2}H}{\partial{\bf \gamma}^{2}})
\ {\bf J }\ ({\bf \nabla}
\rho_{0}^{W})d\qj
\ +\  \int ({\bf \nabla}\rho_{0}^{W})^{T} d\qj\right. \nonumber \\
& & \times \sum_{(l_{1}+l_{2})>1, \ odd}\left.
\frac{(\hbar/2i)^{(l_{1}+l_{2}-1)}}{l_{1}!l_{2}!}{\bf \nabla}
\frac{\partial^{(l_{1}+l_{2})}V(q_{1},q_{2})}{\partial q_{1}^{l_{1}}\partial q_{2}^{l_{2}}}
\frac{\partial^{(l_{1}+l_{2})} \rho^{W}_{0}}{\partial p_{1}^{l_{1}}\partial
p_{2}^{l_{2}}}\right],
\label{lambdaq0}
\eeq
where, obviously, the first term on the right hand side of
Eq. (\ref{lambdaq0}) corresponds to the contribution from the classical Poisson
bracket, and all other terms represent  quantum ``corrections''.

Some aspects of the distribution function strategy outlined above have
been applied to the Arnold-cat map model \cite{gu,arjendu1}, where the
stretching and contraction  mechanism is uniform over the entire phase
space.  However, rich  transient behavior of finite-time Lyapunov
exponents, as implied in our derivation of the explicit state dependence
of $\lambda_{2q}(0)$, has not been explored thus
far.   The next section is devoted to both analytical and numerical
studies on this subject.

\section{Short-time Correspondence }
\label{s2-3}

To examine the QCC in short-time dynamics it is useful to consider the classical analog of $\lambda_{2q}(0)$,
i.e., $\lambda_{2c}(0)$ as the  extreme short
time limit of classical finite Lyapunov exponents. Using Eqs.  (\ref{Mt}), (\ref{m0}), and
(\ref{chi2t}),  we have
\beq
\lim_{t\rightarrow 0} \frac{d\chi_{2c}^{2}}{dt}=
\frac{2}{\int\rho_{0}^{2}({\bf \gamma)}d{\bf \gamma}}\int ({\bf \nabla}
\rho_{0})^{T}(\frac{\partial^{2}H}{\partial{\bf \gamma}^{2}}) {\bf J
}({\bf \nabla} \rho_{0})d\qj.
\label{t0}
\eeq
Substituting Eq. (\ref{t0}) into Eq. (\ref{fdefine}) gives the zero time
limit of $\lambda_{2c}(t)$,
\beq
\lambda_{2c}(0)=\frac{\int ({\bf \nabla}
\rho_{0})^{T}(\frac{\partial^{2}H}{\partial{\bf \gamma}^{2}}) {\bf J}({\bf
\nabla}\rho_{0})d\qj}{\int |{\bf \nabla}\rho_{0}|^{2}d\qj}.
\label{lambda0}
\eeq
Here, $\lambda_{2c}(0)$ is seen to be the average of
$\partial^{2}H/\partial\qj^{2}$ weighted by gradients of the initial
distribution function. Not surprisingly, Eq. (\ref{lambda0}) resembles the first term on the right hand side of
Eq. (\ref{lambdaq0}).   As $\lambda_{2c}(0)$  reflects an ensemble average of
instantaneous density fluctuations, it depends strongly on the shape and
location of the initial classical distribution function. Hence, as in the quantum case, there exists very rich
transient  behavior in the time development of  phase space structure, an
interesting feature that has often been ignored in previous QCC studies.

To consider quantum effects induced solely by the dynamics suggests that
we choose an initial quantal distribution function that is as classical as
possible so that differences between classical and quantum dynamics
evidently arise from the dynamics. This suggests that the initial Wigner
function should be chosen as positive definite, so that it can be
interpreted as a classical probability distribution. It is well known that
for one-dimensional pure state dynamics the only positive-definite Wigner
function is the Gaussian distribution function \cite{hudson,isar}, which
takes the following general form,
\beq
\rho^{W}_{r, \eta, \overline{q}, \overline{p}}=\frac{1}{\pi\hbar}\exp\left[-\frac{2\eta^{2}}{\hbar^{2}}(p-\overline{p})^{2}-
\frac{(q-\overline{q})^{2}}{2\eta^{2}(1-r^{2})}+
\frac{2r}{\hbar(1-r^{2})^{1/2}}(q-\overline{q})(p-\overline{p})\right],
\label{Wigner}
\eeq
 where $r$, $\eta$, $\overline{q}$, $\overline{p}$ are parameters and where
\begin{eqnarray}
&& \langle q\rangle  =  \overline{q},\
\langle p\rangle  = \overline{p},\
\langle q^{2}-\overline{q}^{2} \rangle  =  \eta^{2}, \nonumber\\
&& \langle p^{2}-\overline{p}^{2}\rangle
 =  \frac{\hbar^{2}}{4\eta^{2}(1-r^{2})},\
\langle pq\rangle-\langle p\rangle \langle q\rangle = \frac{\hbar r}{2(1-r^{2})^{1/2}}.
\end{eqnarray}
This Gaussian form  corresponds to the so called correlated coherent
states \cite{dodonov}, whose coordinate representation is given by
\beq
\Psi(q)=\frac{1}{(2\pi\eta^{2})^{1/4}}\exp\left[-\frac{q^{2}}{4\eta^{2}}(1-\frac{ir}{(1-r^{2})^{1/2}})
+\frac{\alpha q}{\eta}-\frac{1}{2}(\alpha^{2}+|\alpha|^{2})\right],
\eeq
where $\alpha$ is complex constant given by $\overline{q}/(2\eta)+
i\left[\overline{p}\eta/\hbar-\overline{q}r/(2\eta\sqrt{1-r^{2}})\right]$.
In particular, for the case of $r=0$,  $\langle q^{2}-\overline{q}^{2} \rangle
\langle p^{2}-\overline{p}^{2}\rangle=\hbar^{2}/4$, corresponding to
the minimum-uncertainty-product state,
i.e., the coherent state.

For the two-degree-of-freedom system examined below one would choose
two-dimensional Gaussian states. The initial quantal distribution function
$\rho^{W}_{0}$ and classical distribution function $\rho_{0}$  are thus
chosen as the following,
\beq
\rho^{W}_{0}=\rho_{0}=\rho^{W}_{r_{1}, \eta_{1}, \overline{q}_{1}, \overline{p}_{1}}\otimes
\rho^{W}_{r_{2}, \eta_{2}, \overline{q}_{2}, \overline{p}_{2}}.
\label{initialrho}
\eeq
Substituting this initial state into Eq. (\ref{lambda0}) and approximating
the average of the derivatives of $V(q_{1}, q_{2})$ as the derivative
evaluated at the centroid of the Gaussian distribution, denoted
$\partial ^{2} V( \overline{q}_{1}, \overline{q}_{2})/\partial
\overline{q}_{i}^{2}$, one gets
\begin{eqnarray}
\lambda_{2c}(0)&=&
\frac{1}{\sum_{i=1,2}\left[2\eta_{i}^{2}
+\frac{\hbar^{2}}{2(1-r_{i}^{2})\eta_{i}^{2}}\right]}
\sum_{i=1,2}\left[\frac{\hbar r_{i}}{(1-r_{i}^{2})^{1/2}}\left(1-\frac{\partial ^{2} V( \overline{q}_{1},
\overline{q}_{2})}{\partial \overline{q}_{i}^{2}}\right)\right].
\label{lambda0theory}
\end{eqnarray}
It is seen that $\lambda_{2c}(0)$ depends strongly on both the shape
parameters $r_{1}, r_{2}$ and the phase space locations of  the initial
distribution function. Two particular situations are worthy of note.
First, for the minimum-uncertainty-product state ($r_{1}=r_{2}=0$), i.e.,
the two-dimensional coherent state that is commonly used as initial states
in QCC studies, Eq. (\ref{lambda0theory}) gives  $\lambda_{2c}(0)=0$.
Second, when $r_{i}\cdot(1-\partial^{2}V(\overline{q}_{1},
\overline{q}_{2})/\partial \overline{q}_{i}^{2})<0$ for either $i=1$ or
$i=2$, $\lambda_{2c}(0)$ can be negative. That is,  for appropriate shape
parameters $r_{1}, r_{2}$ and central coordinates $\overline{q}_{1},
\overline{q}_{2}$, the contraction mechanism associated with chaotic
dynamics may initially dominate over the stretching mechanism, giving rise
to a reduction in phase space structure.

The quantum analog of this transient behavior can be examined  by
considering $\lambda_{2q}(t)$ in a similar fashion.   Specifically,
substituting the initial positive-definte Wigner function
(\ref{initialrho}) into Eq. (\ref{lambdaq0}), keeping the leading order
quantum correction term in the quantum Liouville equation, one obtains
\begin{eqnarray}
\lambda_{2q}(0)&=&\lambda_{2c}(0)
-\frac{1}{16\pi\int|{\bf \nabla}\rho^{W}_{0}|^{2}d\qj}\frac{\partial^{4} V}{\partial
\overline{q}_{1}^{2}\partial \overline{q}_{2}^{2}}
\left[\frac{r_{2}}{(1-r_{2}^{2})^{1/2}}
\int dq_{1} dp_{1}|{\bf \nabla}\rho^{W}_{r_{1}, \eta_{1}, \overline{q}_{1}, \overline{p}_{1}
}|^{2}\right.  \nonumber \\
&& + \left. \frac{r_{1}}{(1-r_{1}^{2})^{1/2}}
\int dq_{2}dp_{2}
|{\bf \nabla}\rho^{W}_{r_{2}, \eta_{2}, \overline{q}_{2}, \overline{p}_{2}}|^{2}\right].
\label{lambdaq02}
\end{eqnarray}
Evidently, initial states with $r_{1}=r_{2}=0$ give $\lambda_{2c}(0)=
\lambda_{2q}(0)=0$. Thus, in this sense, the coherent state is the most
classical state of the correlated coherent states;  other types of initial
states have a leading order quantum effect proportional to $\partial^{4}
V/\partial\overline{q}_{1}^{2}\partial \overline{q}_{2}^{2} $.    After
carrying out the integrals in Eq. (\ref{lambdaq02}) for $\rho^{W}_{r,
\eta, \overline{q}, \overline{p}}$ given by Eq. (\ref{Wigner}), one sees
that $[\lambda_{2c}(0)-\lambda_{2q}(0)]$ is proportional to the first
power of $\hbar$. Therefore, for relatively large $\hbar$, depending upon
the sign of $r_{1}$ and $r_{2}$, $\lambda_{2q}(t)$ can be significantly
larger or smaller than $\lambda_{2c}(t)$ at early times. In addition, if
there is no quartic term in the potential, i.e., $\partial^{4}
V/\partial\overline{q}_{1}^{2}\partial \overline{q}_{2}^{2}=0$ (e.g., in
the Henon-Heiles model), Eq. (\ref{lambdaq02}) shows that
$[\lambda_{2c}(0)-\lambda_{2q}(0)]$ is given by smaller terms that are
proportional to higher powers of $\hbar$.

\begin{figure}[ht]
\begin{center}
\epsfig{file=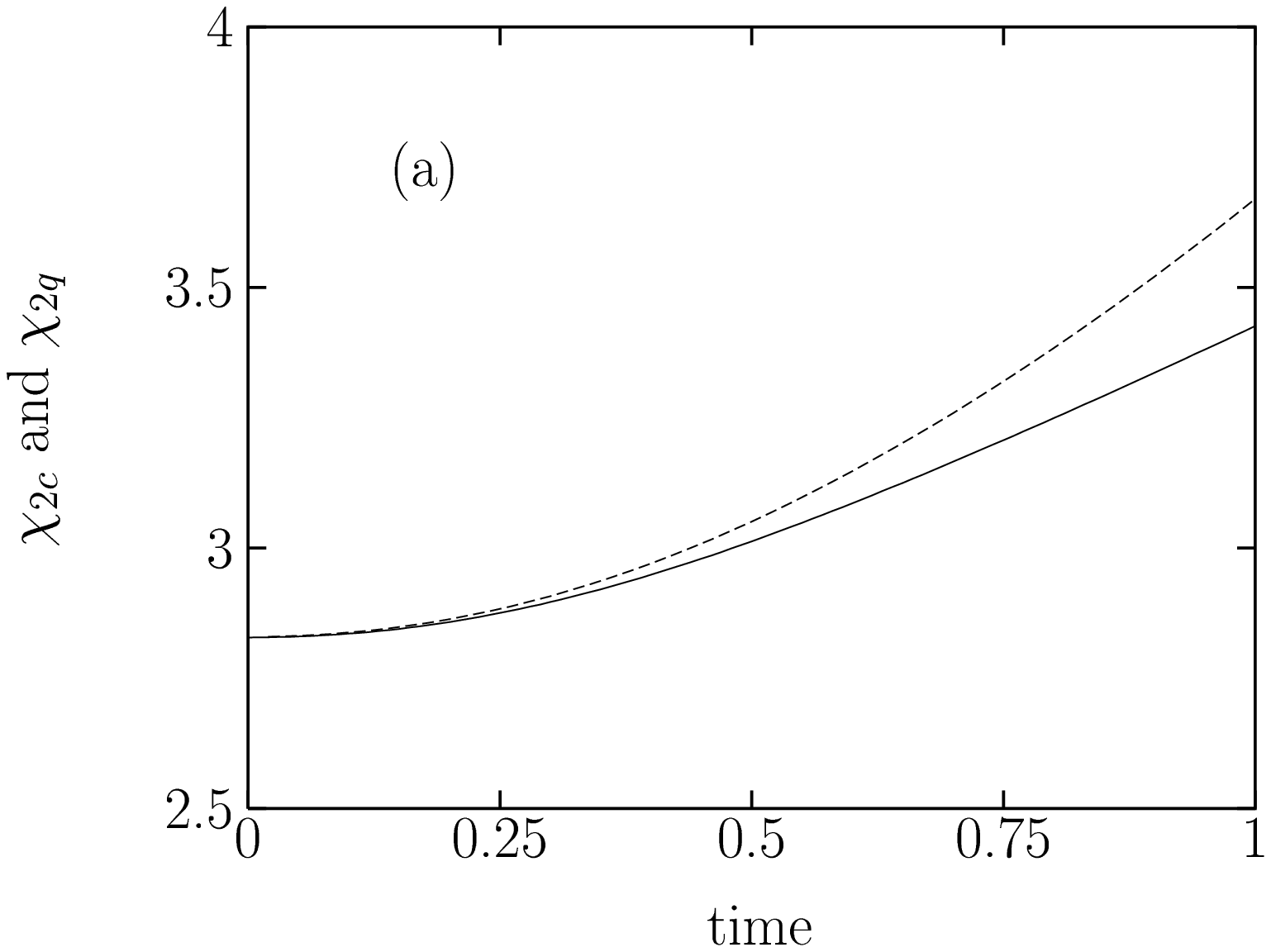,width=8.5cm}

\epsfig{file=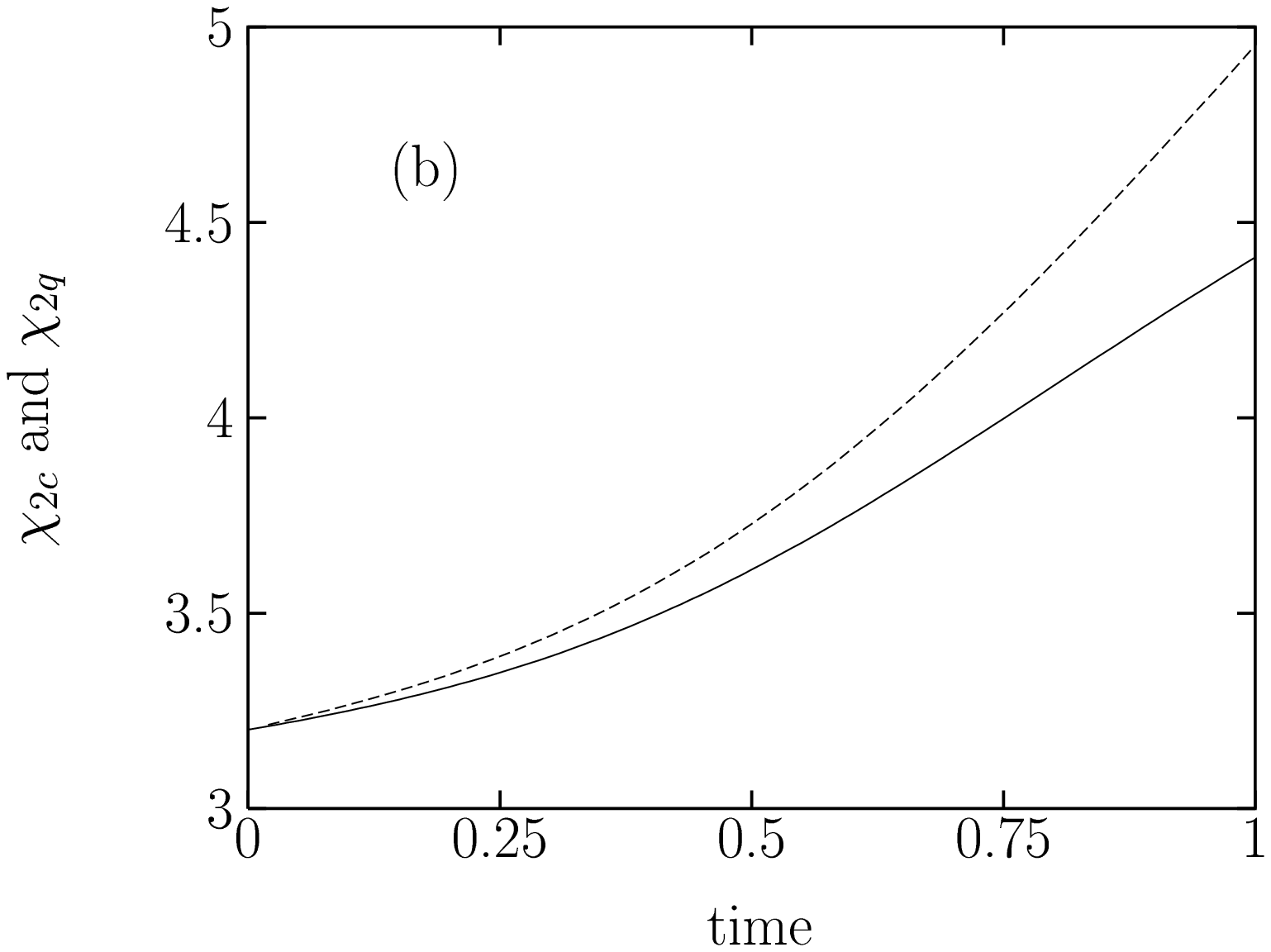,width=8.5cm}

\epsfig{file=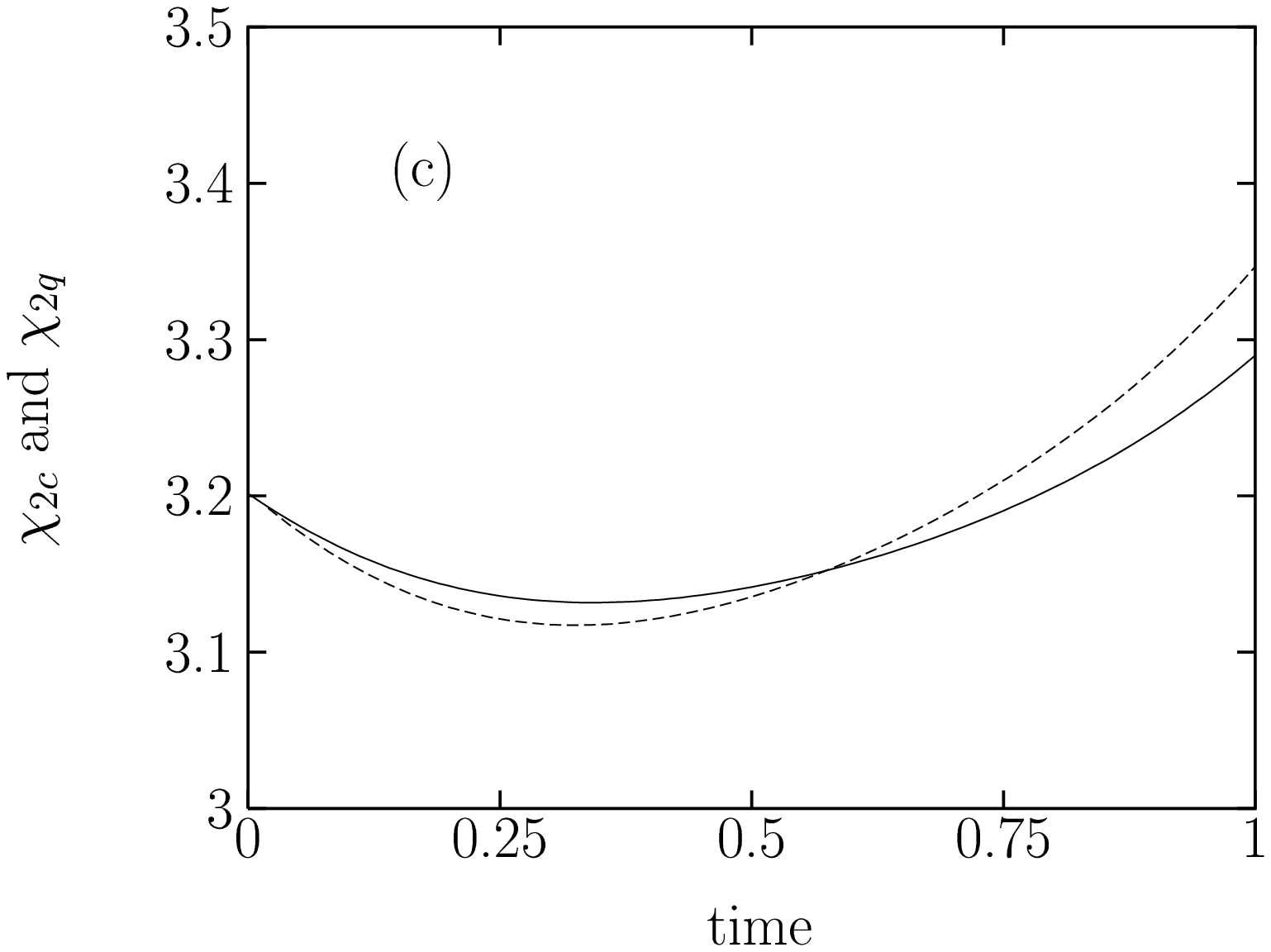,width=8.5cm}
\end{center}
\caption{Time dependence of $\chi_{2c}(t)$ (dashed line)
and $\chi_{2q}(t)$ (solid line) for three initial Gaussian distributions
($\hbar=0.5$).
The three panels correspond to
       (a) $r_{1}=r_{2}=0$,  (b) $r_{1}=r_{2}=0.6$, (c) $r_{1}=r_{2}=-0.6$.
             Note that
the initial slope of  these solid and dashed lines
are zero, positive, and negative in (a), (b) and
(c), respectively.   All  variables
 are in dimensionless units.
}
\label{Fig2-1}
\end{figure}

Thus far we have only examined correspondence between $\lambda_{2c}(t)$
and $\lambda_{2q}(t)$ at $t=0$. For nonzero times, one can utilize
numerical methods to compare these quantities. For example, consider a
coupled quartic oscillator system given by Eq. (\ref{hamiltonian2}) with
$\alpha=1.0$ and $\beta=0.01$, and for $\hbar=0.5$, $0.05$, and $0.005$.
Classical calculations are done by Monte-Carlo methods based on Eq.
(\ref{chi2t}), and quantum calculations use the FFT split operator
technique \cite{fft}. For each case we examine three sets of initial
distributions, i.e., (a) $ r_{1}=r_{2}=0$, (b) $ r_{1}=r_{2}=0.6$, and (c)
$r_{1}=r_{2}=-0.6$. For all three cases
$\eta_{1}=\eta_{2}$=$\sqrt{\hbar/2}$, and the centroid of the initial
state is fixed at $\overline{q}_{1}=0.40$, $\overline{q}_{2}=0.60$,
 $\overline{p}_{1}=0.50$, $\overline{p}_{2}=0.414$.
Note that  this initial location of the Gaussian distribution
gives $(1-\partial^{2}{V}/\partial\overline{q}_{1}^{2})> 0$
and $(1-\partial^{2}{V}/\partial\overline{q}_{2}^{2})> 0
$, a fact that is relevant to the discussions below.

\begin{figure}[ht]
\begin{center}
\epsfig{file=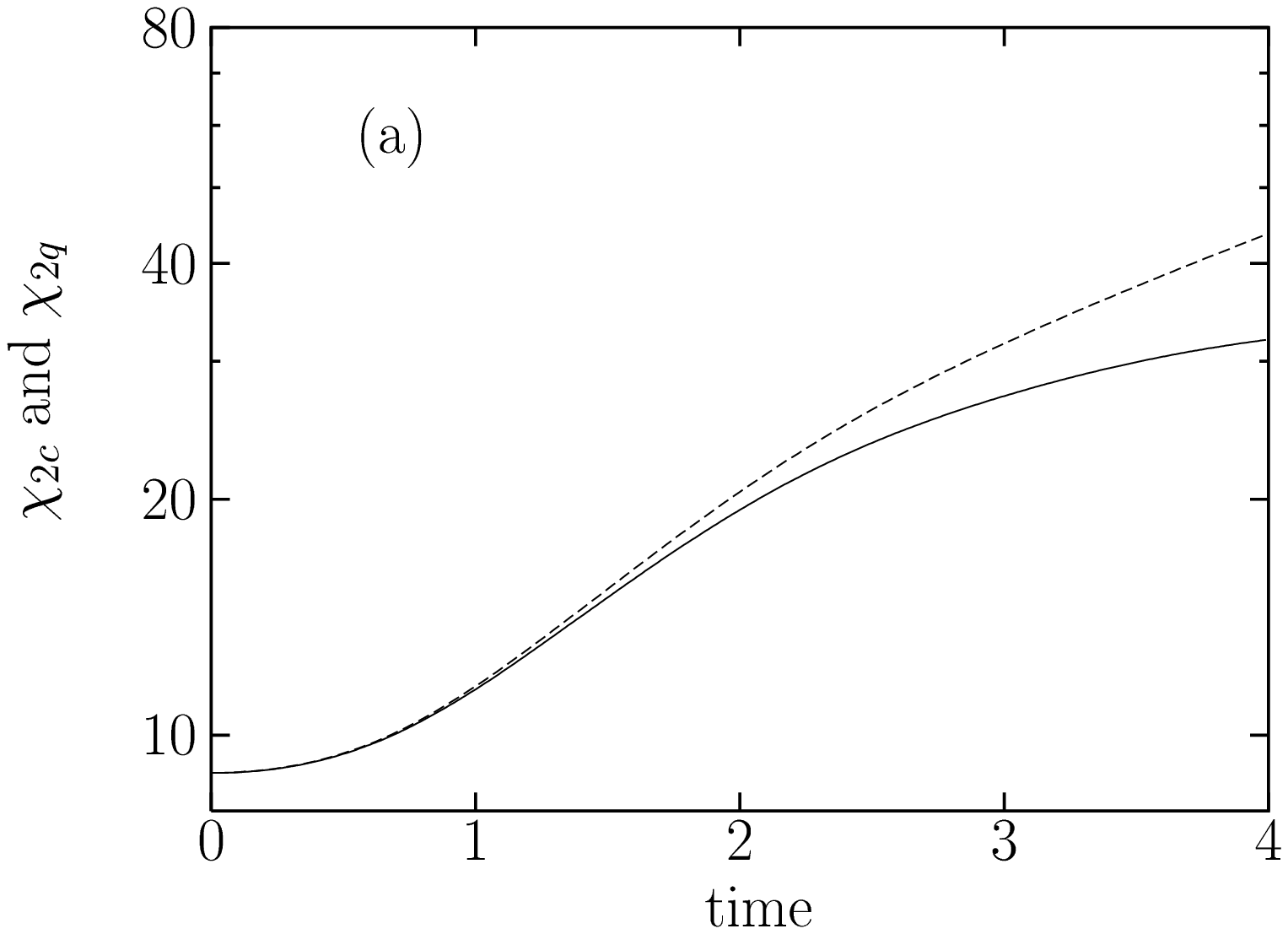,width=8.5cm}

\epsfig{file=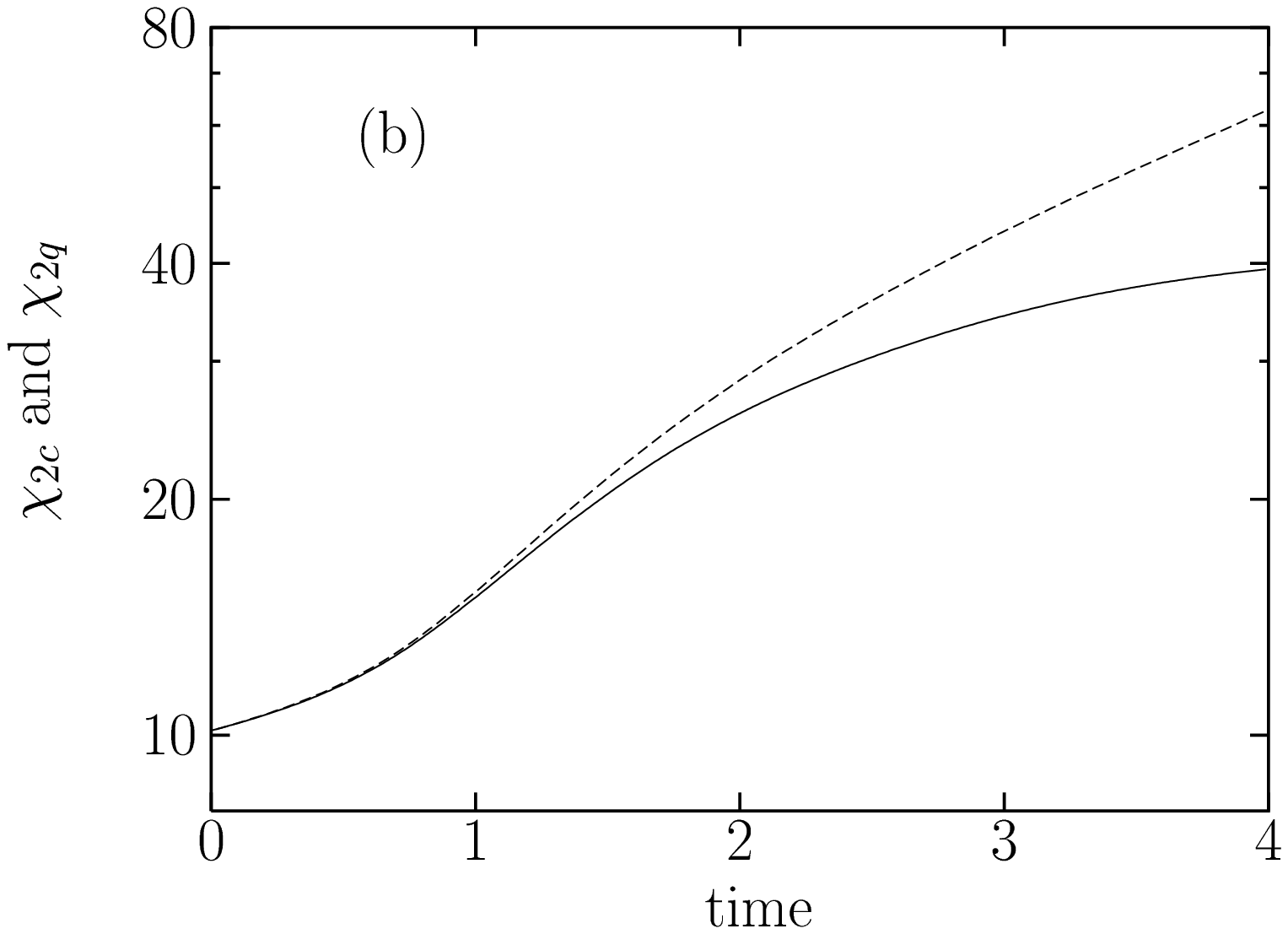,width=8.5cm}

\epsfig{file=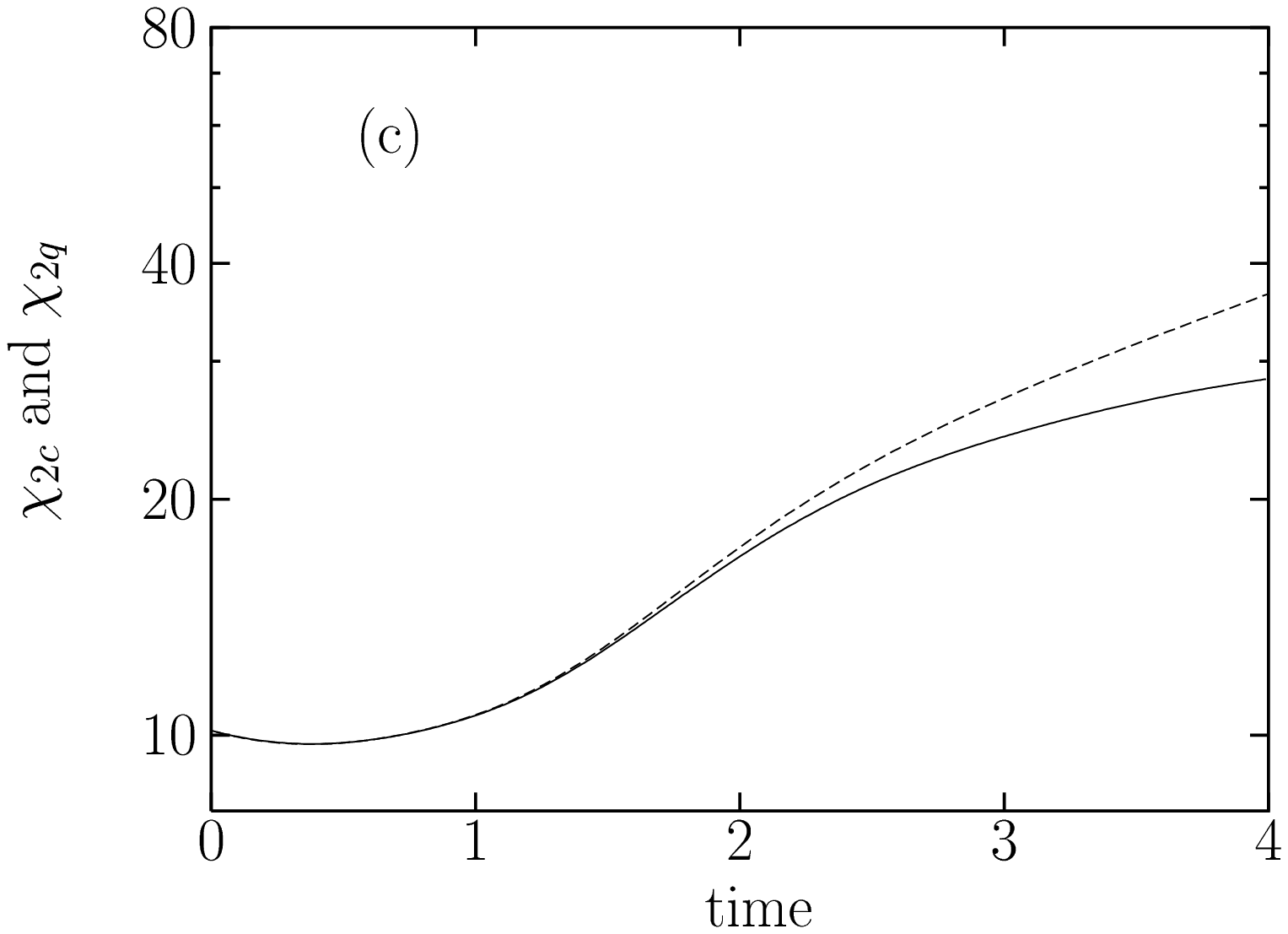,width=8.5cm}

\end{center}
\caption{Same as Fig. \ref{Fig2-1} except $\hbar=0.05$ and that
both $\chi_{2c}(t)$ and $\chi_{2q}(t)$
are plotted in the logarithmic scale.
The average slope of the curves (up to a certain time)
is indicative of the finite-time Lyapunov exponents $\lambda_{2c}(t)$ and
$\lambda_{2q}(t)$.
The break regime of QCC
is clearly
between
$t=1.0$ and
$ t=1.5$. All  variables
 are in dimensionless units.
}
\label{Fig2-2}
\end{figure}

\begin{figure}[ht]

\begin{center}

\epsfig{file=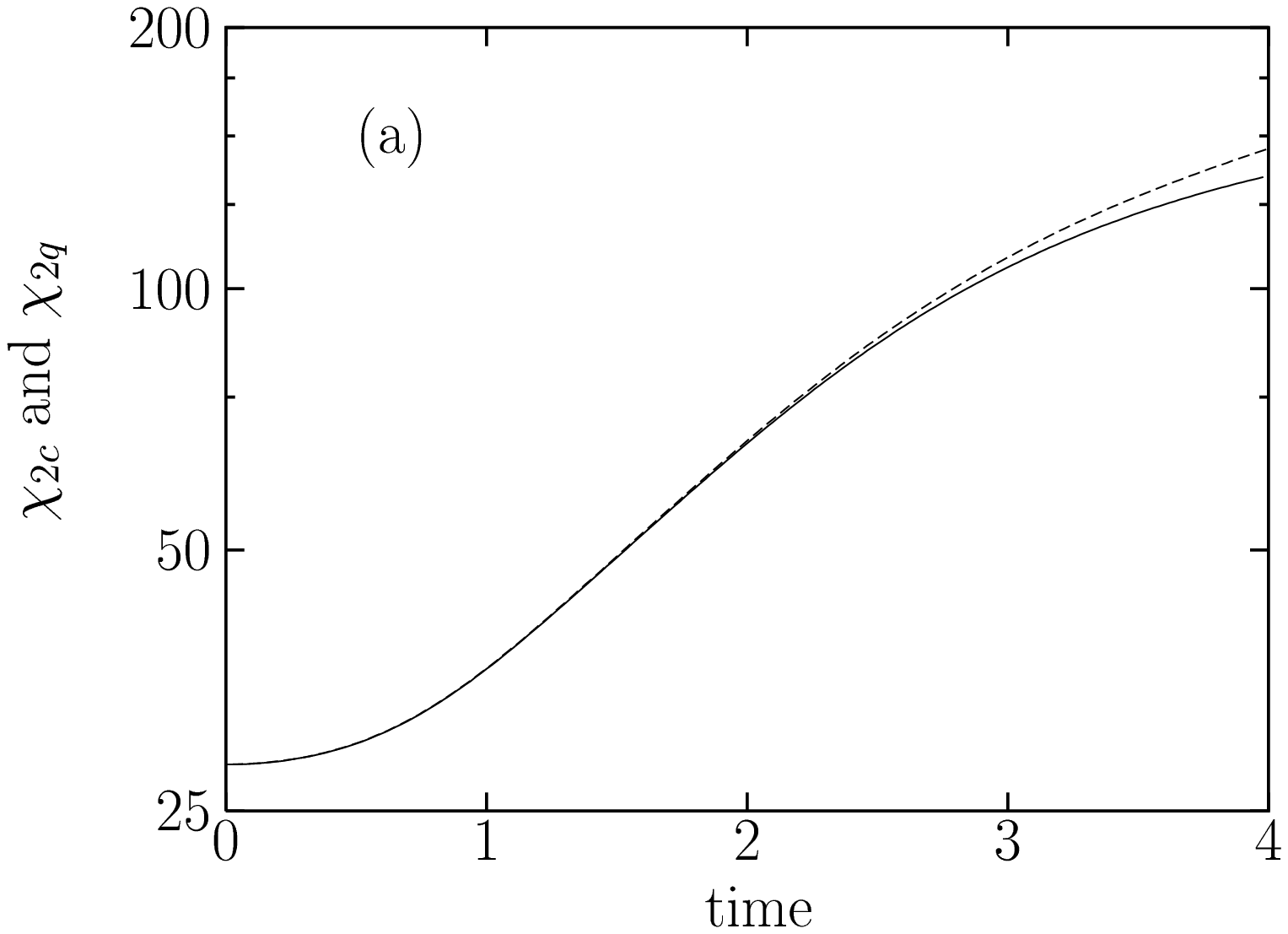,width=8.5cm}

\epsfig{file=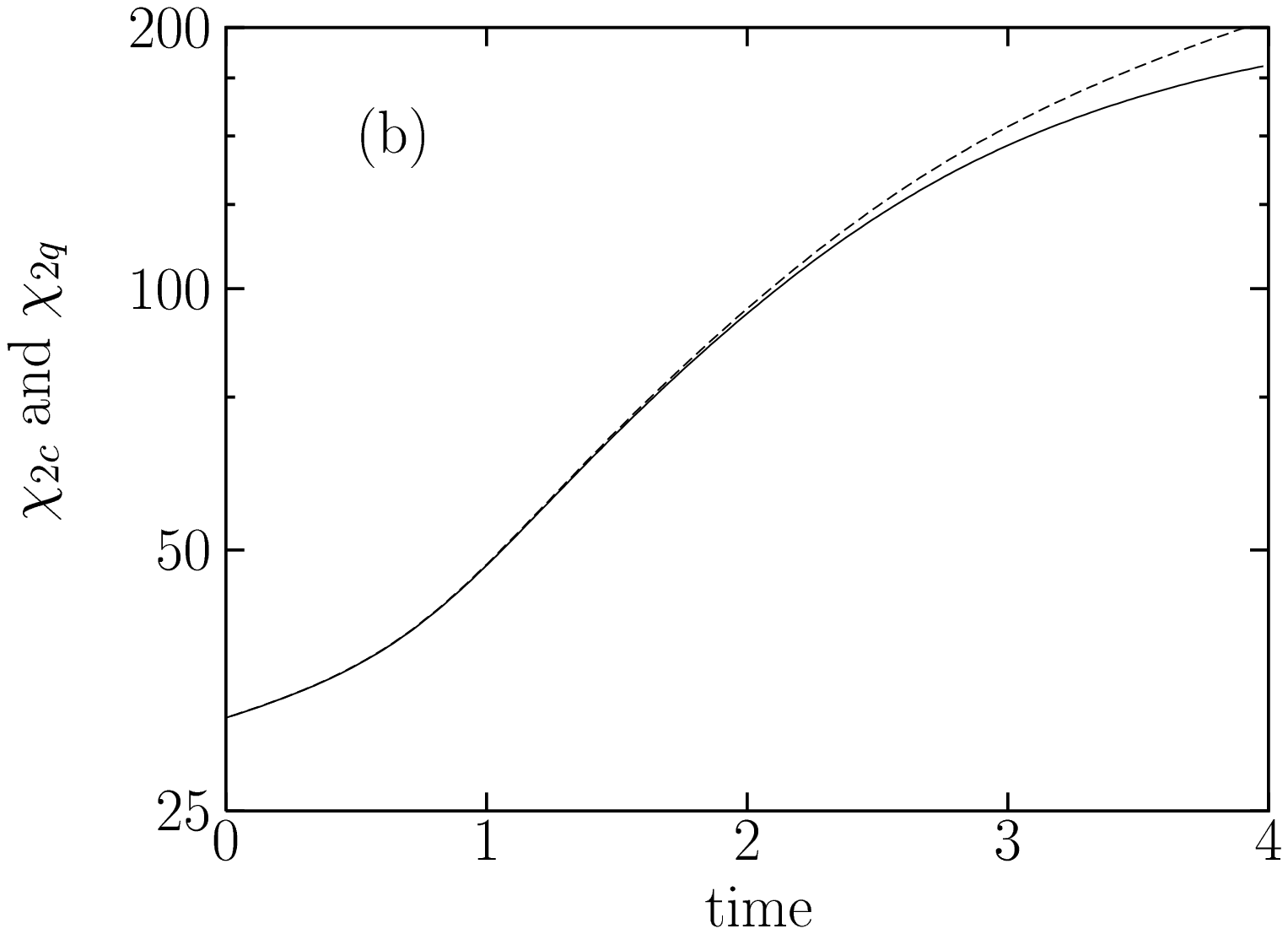,width=8.5cm}

\epsfig{file=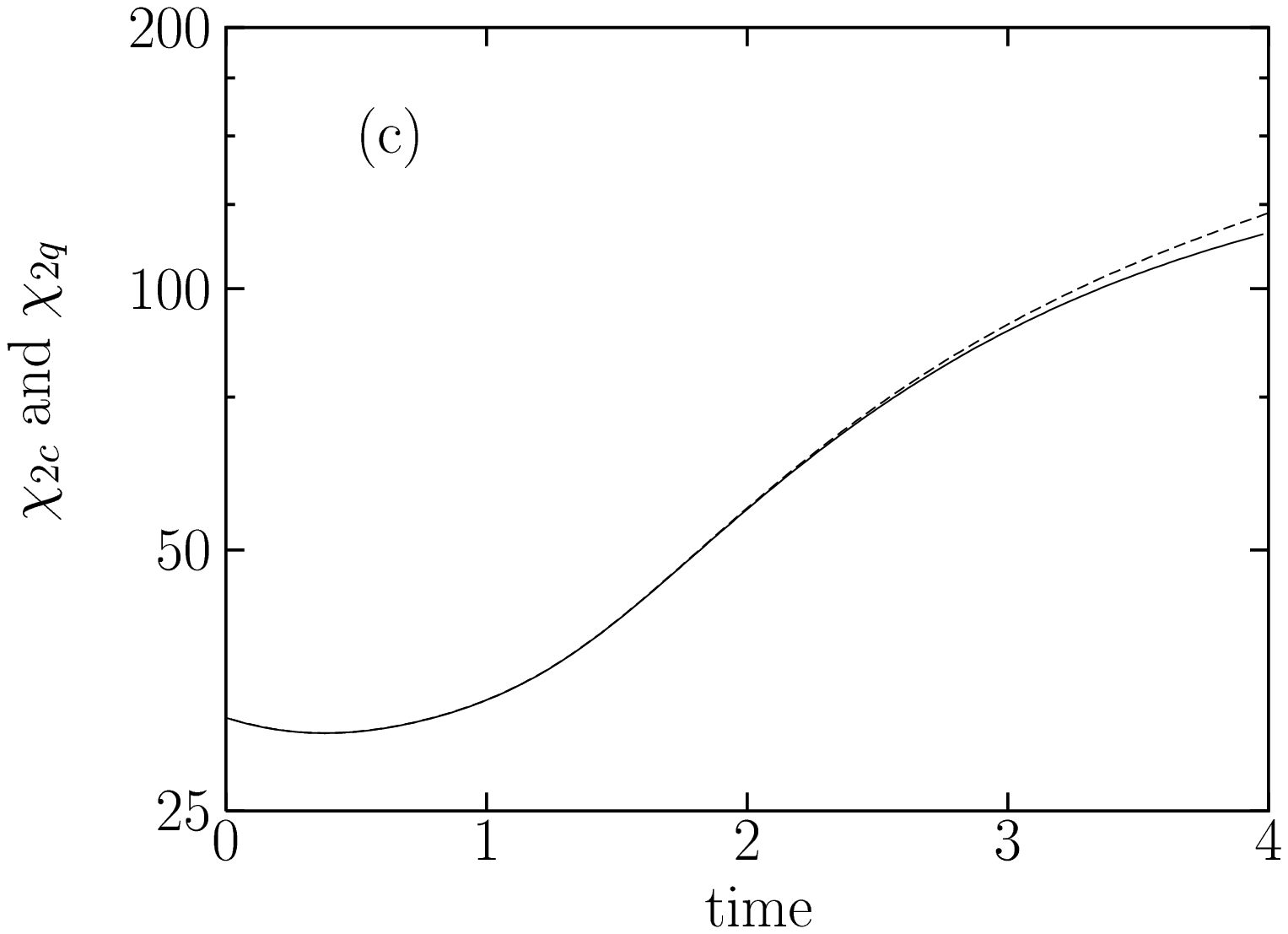,width=8.5cm}

\end{center}
\caption{Same as Fig. \ref{Fig2-2} except $\hbar=0.005$. The break regime of QCC is
clearly between
 $t=2.0$ and
  $t=2.5$.  For $0.5<t<2.5$ and for all three initial conditions
  in (a), (b), and (c), there is  an
    exponential
     increase of phase space structure for both classical and quantum dynamics.
All  variables
 are in dimensionless units.
}
 \label{Fig2-3}
 \end{figure}

Figures \ref{Fig2-1}-\ref{Fig2-3} display comparisons between
$\chi_{2c}(t)$ and $\chi_{2q}(t)$. Consider Fig. \ref{Fig2-1} for the case
of $\hbar=0.5$. Here $\partial^{4} V/\partial\overline{q}_{1}^{2}\partial
\overline{q}_{2}^{2}=2.0$, and one expects that
 $\lambda_{2q}(0)$ can deviate considerably from
$\lambda_{2c}(0)$ for nonzero $r_{1}$ and  $r_{2}$. Further, the
expectation [from Eq. (\ref{lambdaq02})] is that the initial quantum
correction should be positive when  $r_{1},r_{2}$ are negative, and
negative when $r_{1},r_{2}$ are positive. All these predictions are
confirmed nicely in our computations. In particular: (i) in Fig.
\ref{Fig2-1}a ($r_{1}=r_{2}=0$), both $\chi_{2c}(t)$ and $\chi_{2q}(t)$
assume an initial zero slope; (ii) in Fig. \ref{Fig2-1}b
($r_{1}=r_{2}=0.6$), the positive slope of $\chi_{2q}(t)$ at $t=0$ is seen
to be smaller than that of $\chi_{2c}(t)$;  and (iii) in Fig.
\ref{Fig2-1}c ($r_{1}=r_{2}=-0.6$), it is indeed seen that at very short
times $\chi_{2q}(t)>\chi_{2c}(t)$, both of which have negative initial
slopes. For all three situations,  the quantum effects are so large that
QCC is essentially lost at very short times.

Figure \ref{Fig2-2} shows the corresponding results (with different
abscissa scale) with the value of $\hbar$ decreased by a factor of $10$.
This case is different from that in Fig. \ref{Fig2-1} insofar as excellent
QCC is seen for short times for all three initial conditions. In
particular, in Fig. \ref{Fig2-2}b the time dependence of both
$\chi_{2c}(t)$ and $\chi_{2q}(t)$ is seen to be very close to exponential
for $t\le 1.5$. However, this is partially because the initial slope of
the curves [i.e., $\lambda_{2q}(0)$ and $\lambda_{2c}(0)$] in  Fig.
\ref{Fig2-2}b both happen to be close to the average slope (i.e.,
finite-time Lyapunov exponents) for longer times (e.g., $t\sim 4$), In
Fig. \ref{Fig2-2}c  the transient behavior of the finite-time Lyapunov
exponents assumes a completely different nature, i.e.,  the increase of
both $\chi_{2c}(t)$ and $\chi_{2q}(t)$ is significantly suppressed for
times up to $t=0.5$. Variations in the finite-time Lyapunov exponents can
be further seen by comparing $\chi_{2c}(t=4.0)$ and $\chi_{2q} (t=4.0) $
in Fig. \ref{Fig2-2}b to those in Fig. \ref{Fig2-2}a and Fig.
\ref{Fig2-2}c: they can differ by a factor as large as $1.5$. Note also
that Fig. \ref{Fig2-2} shows that the break regime of QCC (i.e., the time
when classical and quantum dynamics no longer agree) is between $t=1.0$
and $t=1.5$, which is on the order of one average period of motion. With
increasing time, the agreement between $\chi_{2q}(t)$ and $\chi_{2c}(t)$
in Fig. \ref{Fig2-2} worsens: the classical phase space structure is seen
to increase exponentially on the average, whereas there is no clear sign
of a similar exponential increase in the quantum distribution dynamics.

As an aside we note that since the structure of quantal distribution
functions may determine the early-time decoherence rate if the quantum
system is open \cite{arjendu2}, these variations in finite-time Lyapunov
exponents imply that even in classically strongly chaotic systems it is
still possible to pick out some initial states which are relatively robust
to decoherence effects.

As shown in Fig. \ref{Fig2-3}, with a further large decrease of $\hbar$,
the break regime of QCC is considerably shifted, to $t=2.0-3.0$. The
transient behavior caused by different initial states still persists for
$t\le 0.5$. After the transient period and before the break time, all
quantal distribution functions  in Figs. \ref{Fig2-3}a-c emanating from
the three different initial conditions are seen to undergo an exponential
increase of structure on the average, in excellent agreement with the
behavior of classical distribution functions.

\section{The Break Regime for Correspondence}
\label{s2-4}

Understanding the QCC break regime is a central problem in the study of
correspondence. From the viewpoint adopted here, i.e., via phase space
distribution functions, the break regime is when the classical
distribution function begins to develop significantly different structure
from the quantal distribution function, i.e., $\chi_{2c}(t)$ begins to
deviate appreciably from $\chi_{2q}(t)$.


It is tempting to try to use the quantum Liouville equation [Eq.
(\ref{qliouville})] to study the breakdown of  QCC  with time. However,
this is not an easy task, since, for nonzero times, one cannot
analytically deal with distribution function dynamics. Rather,  we present
a simple description of the QCC break regime by first re-interpreting
$\chi_{2c}^{2}(t)$ and then comparing it with $\chi_{2q}^{2}(t)$. As will
be quite clear, our description also relates the distribution function
strategy to the trajectory viewpoint of chaos.

It is convenient to  restrict the discussion to the case of the
minimum-uncertainty-product state, although the following considerations
can be readily extended to the most general case.  Further, we focus on
pure state dynamics because mixed state dynamics simply makes the study of
QCC more complicated, without adding new physical insight.

We begin with the Taylor series
 expansion of the stability matrix ${\bf M}[\qj(0), t]$
 around $\qi(0)$; i.e.,
\begin{eqnarray}
{\bf M}_{kl}[\qj(0), t]&=&{\bf M}_{kl}[\qi(0), t] +
\sum_{m}\delta\qj_{m}(0)\frac{\partial {\bf
 M}_{kl}[\qi(0), t]}{\partial
 \qi_{m}(0)}\nonumber \\  &&+
 \frac{1}{2}\sum_{mn}\delta\qj_{m}(0)\delta\qj_{n}(0)\frac{
 \partial ^{2}
 {\bf M}_{
 kl}[\qi(0), t]}{\partial \qi_{m}(0) \partial \qi_{n}(0)}
  +O\left((\delta\qj(0))^{3}\right) \nonumber  \\
 &\equiv& A_{kl} + \sum_{m}\delta\qj_{m}(0)B_{klm} + \sum_{mn}\frac{1}{2}
 \delta\qj_{m}(0)\delta
 \qj_{n}(0)C_{klmn} \nonumber \\
& &  +  O\left((\delta\qj(0))^{3}\right),
 \label{Mtaylor}
 \end{eqnarray}
where we have defined $A_{kl}$, $B_{klm}$, and $ C_{klmn}$,  as
the zeroth, first, and second order derivatives of the stability matrix at $\qi(0)$, respectively.
The summation  indices $m, n,$ etc. run from $1$ to $4$ for a two-degree-of-freedom
system, and $\delta\qj(0)\equiv [\qj(0)-\qi(0)]$.
The time dependence of $A_{kl}$, $B_{klm}$, and $ C_{klmn}$ is
governed by the following set of first order differential equations
together with the canonical equations for classical trajectories:
\begin{eqnarray}
\frac{dA_{kl}}{dt}&=&\sum_{k'l'}J_{kl'}\frac{\partial^{2}H}{\partial\qj_{l'}\partial\qj_{k'}}A_{k'l},\nonumber \\
\frac{dB_{klm}}{dt}&=&\sum_{k'l'}J_{kl'}\left(\sum_{n'}\frac{\partial^{3}H}{\partial\qj_{l'}\partial\qj_{k'}\partial
\qj_{n'}}A_{n'm}A_{k'l}+\frac{\partial^{2}H}{\partial\qj_{l'}\partial\qj_{k'}}B_{k'lm}\right),\nonumber \\
\frac{dC_{klmn}}{dt}&=&
\sum_{k'l'}J_{kl'}\left(\sum_{m'}\frac{\partial^{3}H}{\partial\qj_{l'}\partial\qj_{k'}\partial\qj_{m'}}A_{m'n}
B_{k'lm}+\frac{\partial^{2}H}{\partial\qj_{l'}\partial\qj_{k'}}C_{k'lmn}\right)\nonumber \\
&& + \sum_{k'l'n'}J_{kl'}
\frac{\partial^{3}H}{\partial\qj_{l'}\partial\qj_{k'}\partial
\qj_{n'}}(B_{k'ln}A_{n'm}+B_{n'mn}A_{k'l})
\nonumber \\
&& + \sum_{k'l'n'm'}J_{kl'}
\frac{\partial^{4}H}{\partial\qj_{l'}\partial\qj_{k'}\partial\qj_{m'}\partial\qj_{n'}}A_{k'l}A_{n'm}A_{m'n}.
\label{ABC}
\end{eqnarray}

Consider two initial points in the phase space, $\qj(0)$ and $ \qi(0)$.
They generate two classical trajectories, denoted by $\qj(t)$ and $ \qi(t)$, respectively.
From the trajectory viewpoint, of most interest is the difference  $\delta\qj_{j}(t)\equiv
[\qj_{j}(t)-\qi_{j}(t)]$, which is a differentiable
 function of $\qi(0)$, $\delta\qj(0)$, and $t$.
Fixing $\qi(0)$, and thus $\qi(t)$, gives a reference trajectory.
Consider now the Taylor series
expansion of $\delta\qj_{j}(t)$  in terms of  $\delta\qj(0)$.
Obviously  $\delta\qj_{j}(t)=0$ if $\delta\qj(0)=0$,
$\partial \delta\qj_{j}(t)/\partial \delta\qj_{l}(0)|_{\delta\qj(0)=0}= A_{jl}$,
$ \partial^{2}\delta\qj_{j}(t)/
\partial \delta\qj_{k}(0)
\partial \delta\qj_{l}(0)|_{\delta\qj(0)=0}= B_{jkl}$ , and so on.
In fact, the $n$th order derivative of $\delta\qj_{j}(t)$ with respect to
$\delta\qj(0)$ is simply given by the $(n-1)$th order derivative of the
stability matrix ${\bf M}[\overline{\qj}(0),t]$ with respect to
$\overline{\qj}(0)$. Based on this observation,  one obtains the following
relation:
\begin{eqnarray}
\sum_{l}A_{jl}\delta\qj_{l}(0)&=&\delta\qj_{j}(t)-\frac{1}{2}
\sum_{lk}\delta\qj_{k}(0)\delta\qj_{l}(0)B_{jkl}
\nonumber \\  &&
-  \frac{1}{6}\sum_{jklm}\delta\qj_{k}(0)\delta\qj_{l}(0)\delta\qj_{m}(0)C_{jklm}
 -  O\left((\delta\qj(0))^{4}\right).
\label{Md}
\end{eqnarray}

Consider now the initial classical distribution function corresponding to
the coherent state
\begin{eqnarray}
 \rho_{0}[\qj(0), \qi(0)]& =  &
\left(\frac{1}{4\pi^{2}\sigma_{q_{1}}\sigma_{q_{2}}\sigma_{p_{1}}\sigma_{p_{2}}}\right)
 \exp\left[-
\frac{(q_{1}-\overline
{q}_{1})^{2}}{2\sigma_{q_{1}}^{2}}-
\frac{(q_{2}-\overline{q}_{2}^{2})}{2\sigma_{q_{2}}^{2}}\right.\nonumber \\
& &\left.-\frac{(p_{1}-\overline{p}_{1})^{2}}{2\sigma_{p_{1}
}^{2}
}-\frac{(p_{2}-\overline{p}_{2})^{2}}{2\sigma_{p_{2}}^{2}}\right].
\label{distri}
\end{eqnarray}
After lengthy calculations (see Appendix)
using Eqs. (\ref{chi2t}), (\ref{Mtaylor}), (\ref{Md}) and (\ref{distri}),
we obtain an enlightening expression for $\chi_{2c}^{2}(t)$:
\begin{eqnarray}
\chi_{2c}^{2}(t)&= &\frac{2}{\hbar^{2}}\sum_{j}\left[\langle(\qj_{j}(t))^{2}\rangle_{0}
-\langle\qj_{j}(t)\rangle_{0}^{2}\right] \nonumber \\
&&- \frac{2}{\hbar^{2}}\sum_{jkk'll'}\left(\frac{1}{4}B_{jkl}B_{jk'l'}+\frac{1}{3}A_{jk}C_{jk'll'}\right)
\langle\delta\qj_{k}(0)\delta\qj_{k'}(0)\delta\qj_{l}(0)\delta\qj_{l'}(0)\rangle_{0} \nonumber \\
&&+ \sum_{jkk'll'}\left(\frac{1}{4}B_{jkl}B_{jk'l'}+\frac{1}{4}A_{jk}C_{jk'll'}\right)
\langle\alpha_{k}\alpha_{k'}\delta\qj_{l}(0)\delta\qj_{l'}(0)\rangle_{0}  \nonumber \\
&& + \frac{1}{\hbar^{2}}O\left((\delta\qj(0))^{6}\right).
\label{chi2expansion1}
\end{eqnarray}
Here $\langle\cdot\rangle_{0}$ represents the average over the initial
classical ensemble. Note that the factor $\hbar^{2}$  in the above
equation is due to the fact that
$\sigma_{q_{1}}\sigma_{p_{1}}=\sigma_{q_{2}}\sigma_{p_{2}}=\hbar/2$ for
any classical distribution function  corresponding to a
minimum-uncertainty-product state.

Equation (\ref{chi2expansion1}) provides a quantitative connection between
the development of phase space structure and the  instability
characteristics of classical trajectories. Further, it allows for a closer
examination of when and how classical distribution functions begin to
develop structure that differs from quantal distribution functions.
Consider then, for simplicity, a special case in which the initial
coordinate variances are identical to the initial momentum variances,
i.e., $\sigma_{q_{1}}=\sigma_{p_{1}} =\sigma_{q_{2}}=\sigma_{p_{2}}$.  One
obtains
\beq
\chi_{2c}^{2}(t) = \frac{2}{\hbar^{2}}\sum_{j}\left[\langle(\qj_{j}(t))^{2}\rangle_{0}-
\langle\qj_{j}(t)\rangle_{0}^{2}\right] + f(t) + O(\hbar),
\label{foft}
\eeq
where $f(t)$ is given by
\begin{eqnarray}
f(t) &=& -\sum_{jkk'll'}\left(\frac{1}{2}B_{jkl}B_{jk'l'}+\frac{2}{3}A_{jk}C_{jk'll'}\right) \nonumber \\
&&\times \left[\frac{3}{4}\delta_{kk'll'}+\frac{1}{4}\left[(\delta_{kk'}\delta_{ll'}
 +\delta_{kl'}\delta_{lk'})(1-\delta_{kl})+\delta_{kl}\delta_{k'l'}(1-\delta_{kk'})\right]
 \right] \nonumber
 \\
&& +\sum_{jkk'll'mm'}J_{km}J_{k'm'}\left(B_{jkl}B_{jk'l'}+A_{jk}C_{jk'll'}\right)\nonumber \\
&& \times \left[\frac{3}{4}\delta_{mm'll'}
+\frac{1}{4}\left[(\delta_{mm'}\delta_{ll'}
+\delta_{ml'}\delta_{lm'})(1-\delta_{ml})
+ \delta_{ml}\delta_{m'l'}(1-\delta_{mm'})\right] \right] .
\label{fexp}
\end{eqnarray}

This expression affords new perspectives into QCC. Recall first, as shown
in the previous  section [see  Eq. (\ref{chi2q})], that for pure state
dynamics the quantal measure $\chi_{2q}^{2}(t)$ of phase space structure
can be expressed in terms of the sum of second order moments.  This is
more or less a consequence of quantization: the smallest scale in
coordinates is related to the largest momentum component of the
wavefunction, and vice versa. Clearly, this is in general not the case for
classical distribution functions. Nevertheless,  Eqs.
(\ref{chi2expansion1}) and (\ref{foft}) indicate that, to lowest order,
$\chi_{2c}^{2}(t)$ can still be related to the sum of  second order
statistical moments. Specifically, as shown in Eq. (\ref{foft}), for an
initially symmetric coherent state the first term contributing to
$\chi_{2c}^{2}(t)$ is given by the sum of second order moments divided by
$\hbar^{2}/2$,  which is exactly the same as the result in Eq.
(\ref{chi2q}) for $\chi_{2q}^{2}(t)$. In addition, $\chi_{2c}^{2}(t)$
contains additional  contributions absent in the quantum dynamics. For
example, the leading order correction term $f(t)$ is {\em independent of}
$\hbar$.  As shown by Eq. (\ref{fexp}), this term is determined by the
stability characteristics $ A_{jk}$, $B_{jkm}$, and $C_{jkmn} $ associated
with the trajectory starting from the centroid of the initial Gaussian
distribution.  These results show intriguing similarities and differences
between $\chi_{2c}(t)$ and $\chi_{2q}(t)$.

As a  simple example of Eq. (\ref{foft}), consider a quadratic Hamiltonian
systems such as the harmonic oscillator or the inverted harmonic
oscillator system, where classical and quantum mechanics are expected to
agree. Here the time evolution is a linear canonical transformation in
phase space; hence the stability matrix elements ${\bf
M}_{jk}[\overline{\qj}(0), t]$ do not depend on $\overline{\qj}(0)$ and
$B_{jkl}, C_{jklm}$ and all other higher order derivatives of the
stability matrix with respect to $\overline{\qj}(0)$ are zero. Thus, in
such linear systems, $\chi_{2c}^{2}(t)$ is precisely given by the sum of
some second order moments divided by $\hbar^{2}/2$, in perfect
correspondence with  $\chi_{2q}^{2}(t)$ [see Eq. (\ref{chi2q})].

For the case of chaotic systems,  one can estimate that the stability
matrix increases exponentially, i.e., $A_{kl} \sim \exp(\lambda t)$, with
$\lambda$ being the average exponential increase rate up to time $t$.
Likewise, one expects $B_{klm} \sim \exp(\lambda t)$,  $C_{klmn}\sim
\exp(\lambda t)$, etc. Equation (\ref{fexp}) then suggests that $f(t)\sim
f_{0}\exp(2\lambda t)$. By contrast, the first term on the right hand side
of Eq. (\ref{foft}) is determined  by second-order moments and cannot
increase for all time for bounded systems. Indeed,
 assuming that the characteristic magnitude
 of the second order moments $[\langle(\qj_{j}(t))^{2}\rangle_{0}-
 \langle\qj_{j}(t)\rangle_{0}^{2}]$ $(j=1,2,3,4)$
is given by $\Omega^{2}$, then the first term on the
right hand side of Eq. (\ref{foft}) would be bounded
by $8\Omega^{2}/\hbar^{2}$.   Thus, in  Eq. (\ref{foft}) the $f(t)$ term
 will be comparable to the
preceding  term after a time $t_{b}$,  approximately given by
\begin{eqnarray}
t_{b}=\frac{1}{\lambda}\ln \left[\frac{\sqrt{8\Omega^{2}/f_{0}}}{\hbar}\right].
\label{tb}
\end{eqnarray}
A comparison between Eqs. (\ref{chi2q}) and (\ref{foft}) suggests that
$t_{b}$ corresponds to the time scale after which classical descriptions
of phase space structure no longer agree with quantum results. Thus,
$t_{b}$ can be identified as a logarithmic break time of QCC. This result
is consistent with previous studies on the QCC break time using different
approaches \cite{logtime}. Note that,  since Eq. (\ref{tb}) involves
classical variables only, one can calculate $t_{b}$ without the need for
any quantum calculations.

\begin{figure}[ht]
\begin{center}

\epsfig{file=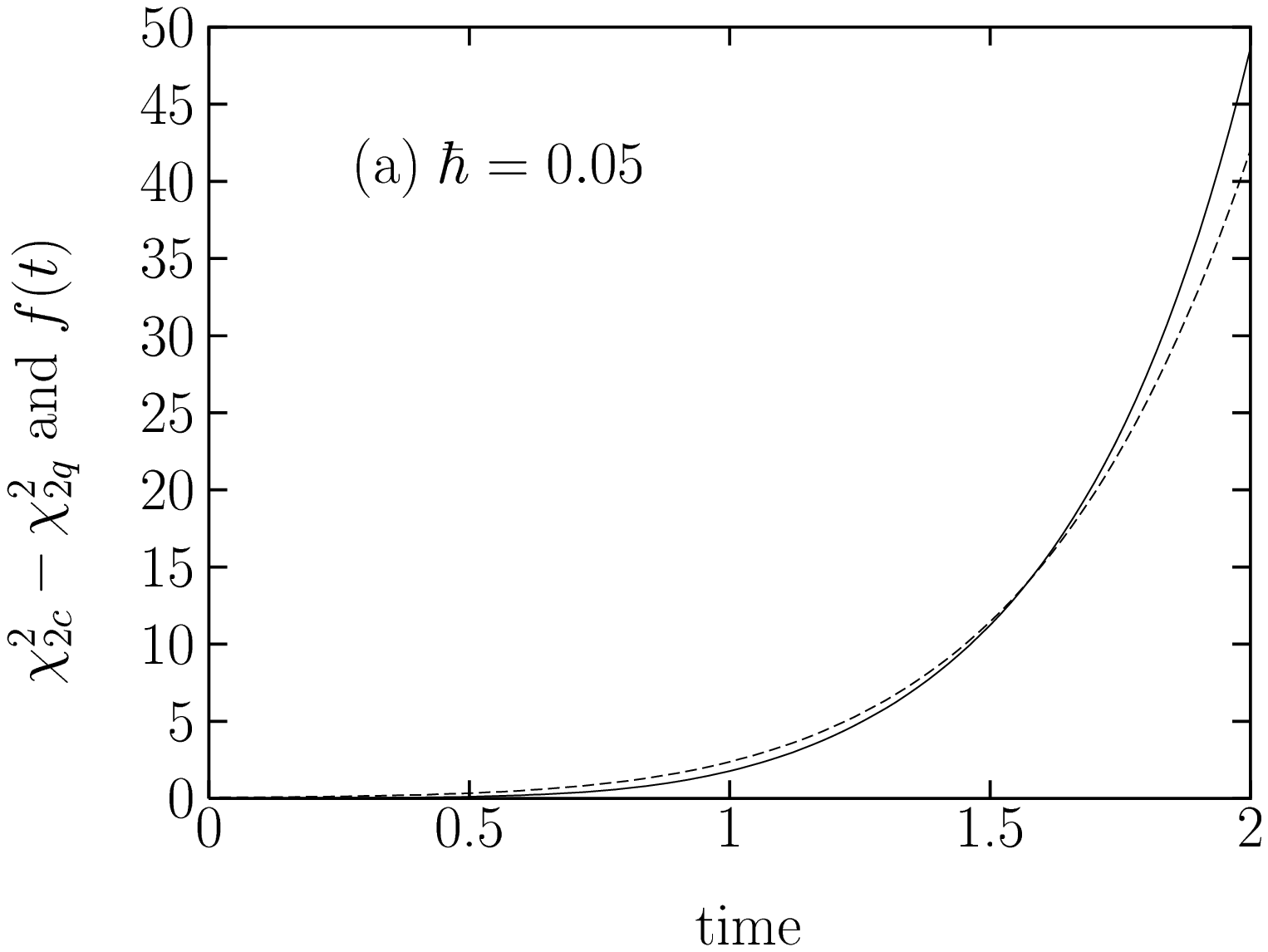,width=8.cm}
\epsfig{file=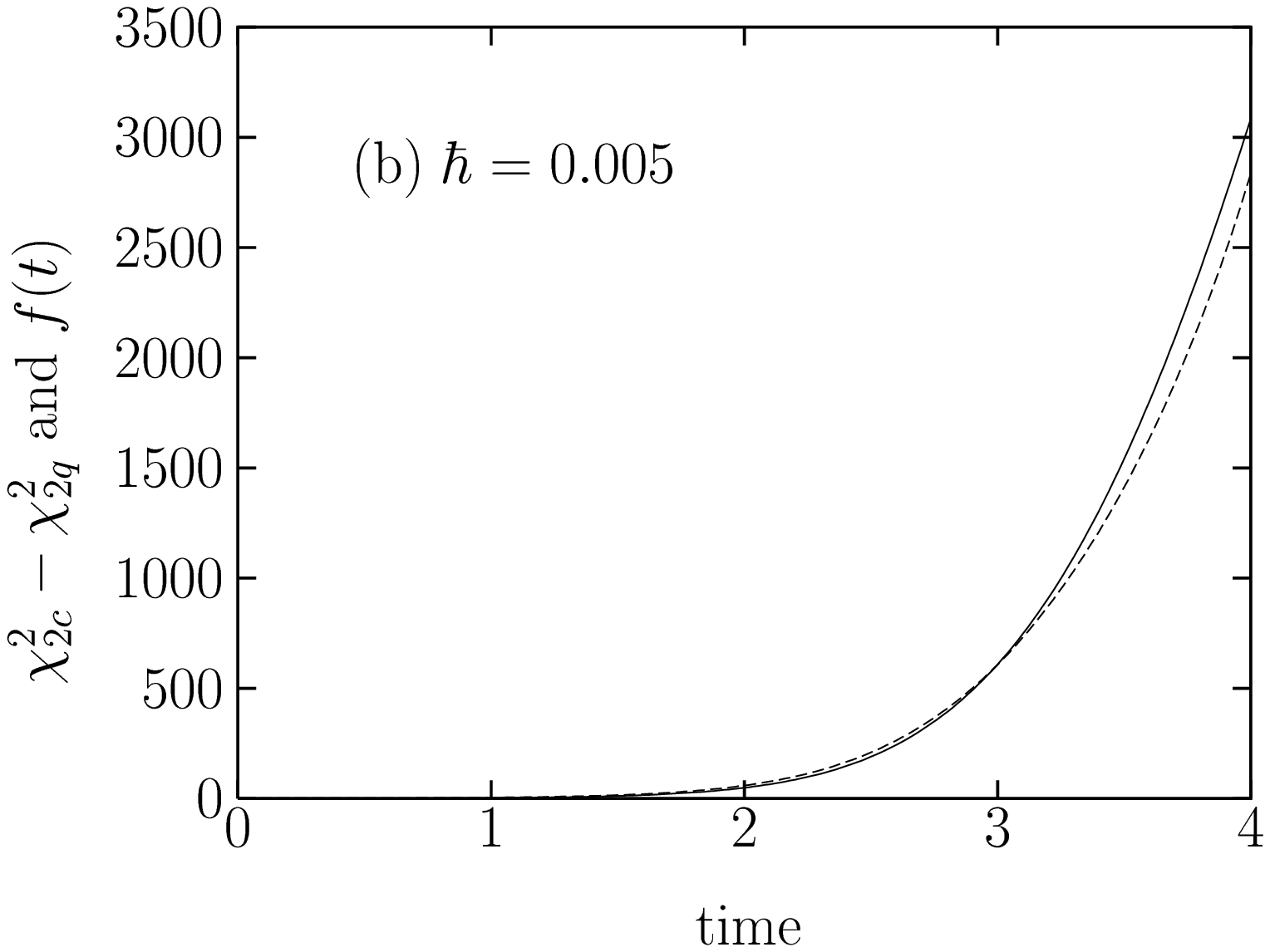,width=8.cm}
\end{center}
\caption{Time dependence of ($\chi_{2c}^{2}-\chi_{2q}^{2})$ compared with
$f(t)$ in our analytical considerations [see Eqs. (\ref{foft}) and (\ref{fexp}) in
the text]. The initial condition is the
minimum-uncertainty-product state
considered in Sec. \ref{s2-3}.
Dashed lines denote $(\chi_{2c}^{2}-\chi_{2q}^{2})$ based on direct calculations
in Sec. \ref{s2-3},
and solid lines denote $f(t)$ obtained by directly integrating Eq. (\ref{fexp}).
$\hbar$ equals 0.05 in (a) and equals 0.005 in (b).
The agreement between $(\chi_{2c}^{2}-\chi_{2q}^{2})$ and $f(t)$
for the QCC break regime is excellent.
All variables are in dimensionless units.
}
\label{Fig2-4}
\end{figure}

The origin of the classical-quantum difference lies in the $f(t)$ term in
Eq. (\ref{foft}). This term  does not have a quantal analog since it
reflects classical phase space structure that is  beyond the resolution
limit of quantal distribution functions \cite{resolution-note}. To
demonstrate the role of $f(t)$ we compare  $[\chi_{2c}^{2}(t)-
\chi_{2q}^{2}(t)]$  from a direct calculation to $f(t)$ given  by Eq.
(\ref{fexp}).  Specifically,  $f(t)$ is obtained from Eq. (\ref{fexp}) by
numerically computing the instability characteristics $A_{jk}$, $B_{jkl}$
and $C_{jklm}$ via Eq. (\ref{ABC}). Two cases with differing values of
$\hbar$ have been examined. Results are shown in Figs. \ref{Fig2-4}a and
\ref{Fig2-4}b that display the comparison between $[\chi_{2c}^{2}(t)-
\chi_{2q}^{2}(t)]$ and $f(t)$ for both cases. The agreement is excellent,
confirming the role of $f(t)$ in determining the break regime of QCC.

Figure \ref{Fig2-4} also shows some tiny discrepancies between
$[\chi_{2c}^{2}(t)- \chi_{2q}^{2}(t)]$ and $f(t)$ (especially at later
times), corresponding to higher order contributions in Eq. (\ref{foft}).
Nevertheless, the physics is still the same; i.e., the sum of $f(t)$ and
these higher order terms  measures very fine phase space structure that is
unresolvable by the quantum distribution function.

\section{Long After the Break Regime}
\label{s2-5}

The  logarithmic break time $t_{b}$ of QCC introduces many issues
regarding the relationship between quantum dynamics and classical
mechanics. For instance, Zurek {\it et al.} \cite{zurek} showed that a
logarithmic break time (somewhat different from that obtained above) can
be counter-intuitively short even for macroscopic objects. This being the
case, our everyday experience in a macroscopic classical world full of
chaotic events seems incompatible with the notion that classical physics
is a large-quantum-number limit of quantum mechanics. Likewise, the
smallness of the QCC break time seems to imply that classical physics
cannot play a role in nonlinear molecular dynamics, contradicting the fact
that classical physics often  works well in many dynamics simulations. To
at least partially resolve these puzzles, this section attemps to explore
the implications of the quantum-classical discrepancy in phase space
structure for ensemble statistics.

\begin{figure}[ht]
    
\vspace{-0.3cm}
\begin{center}
\epsfig{file=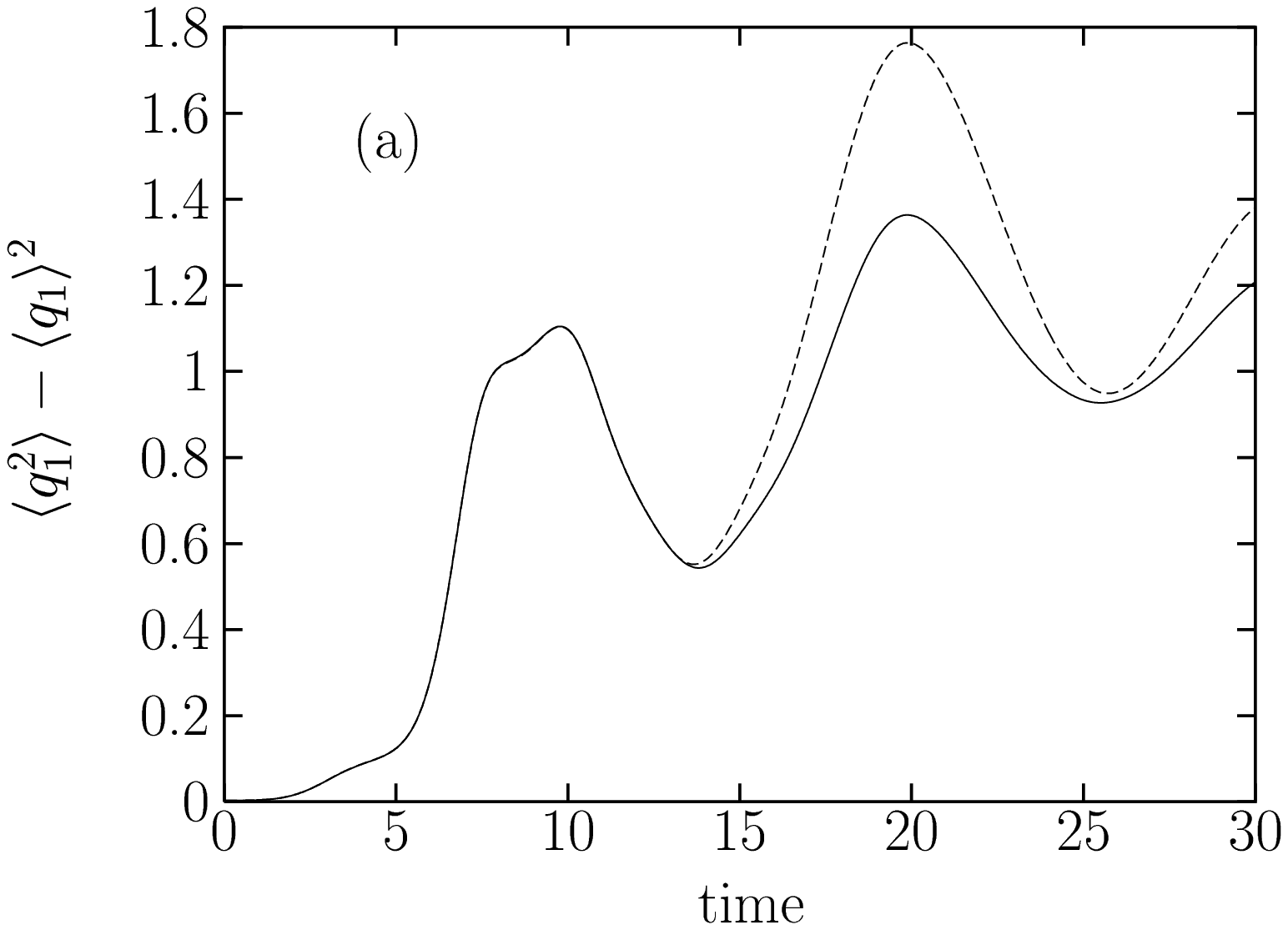,width=8.0cm}
\epsfig{file=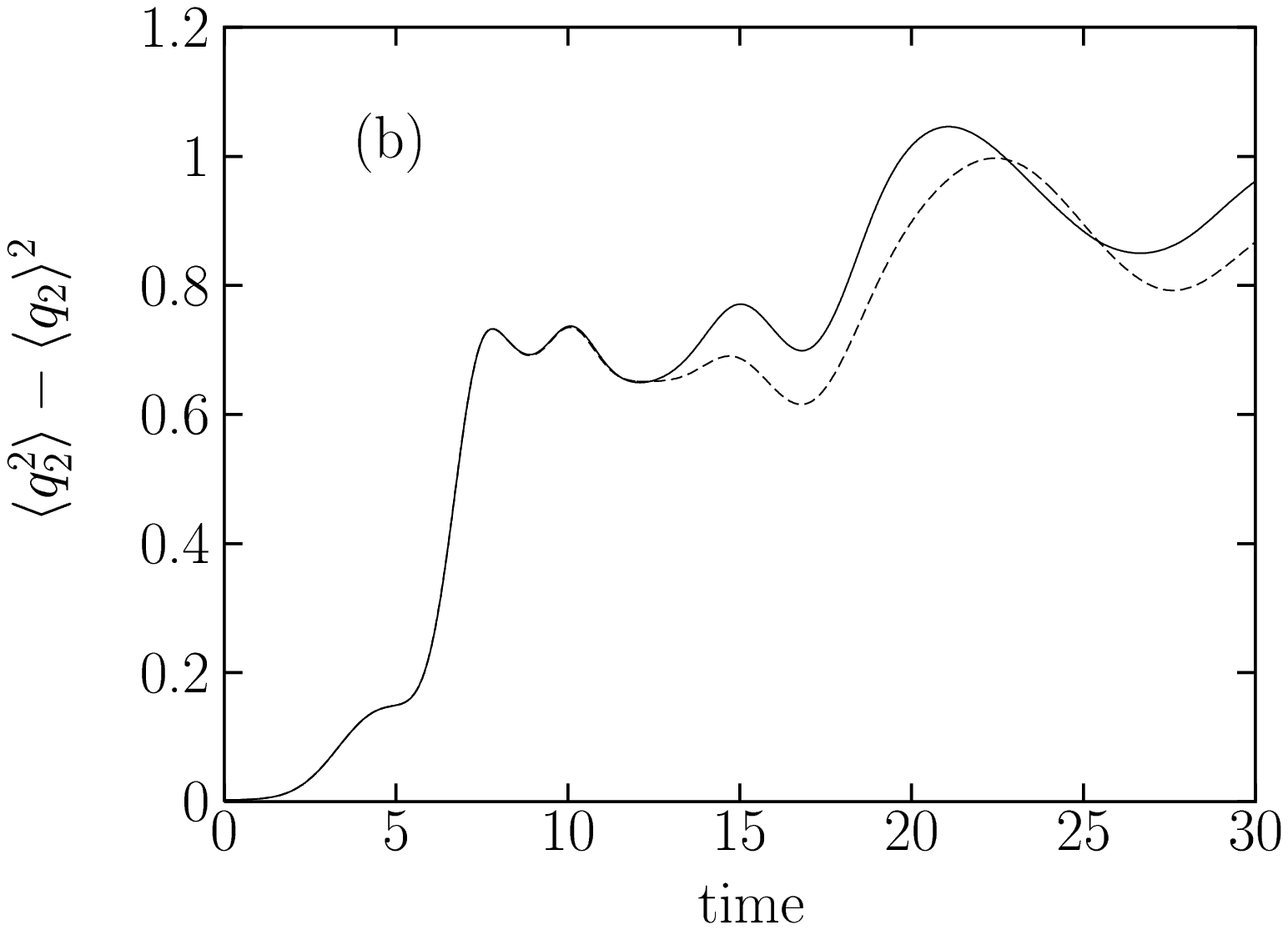,width=8.0cm}

\epsfig{file=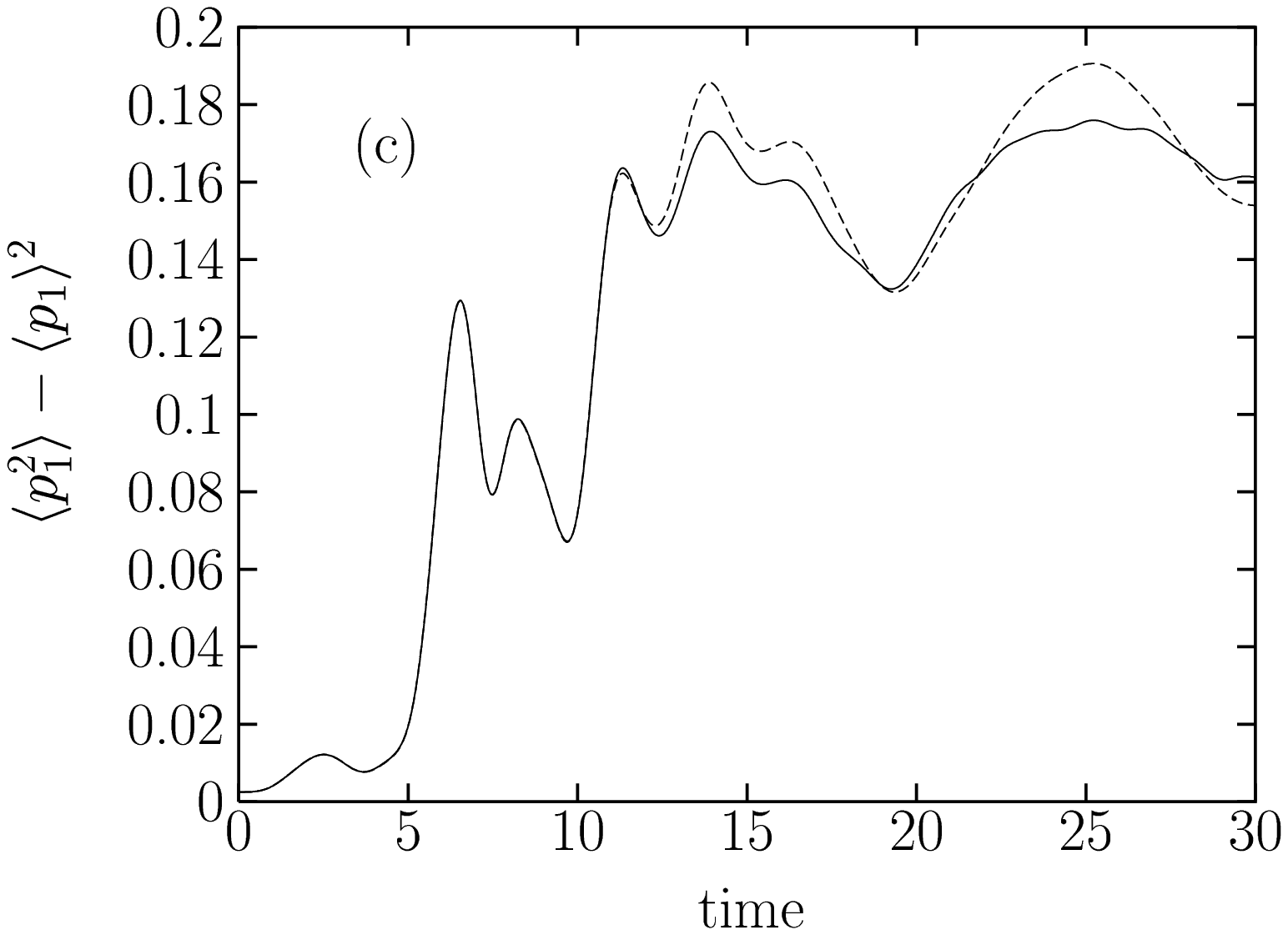,width=8.0cm}
\epsfig{file=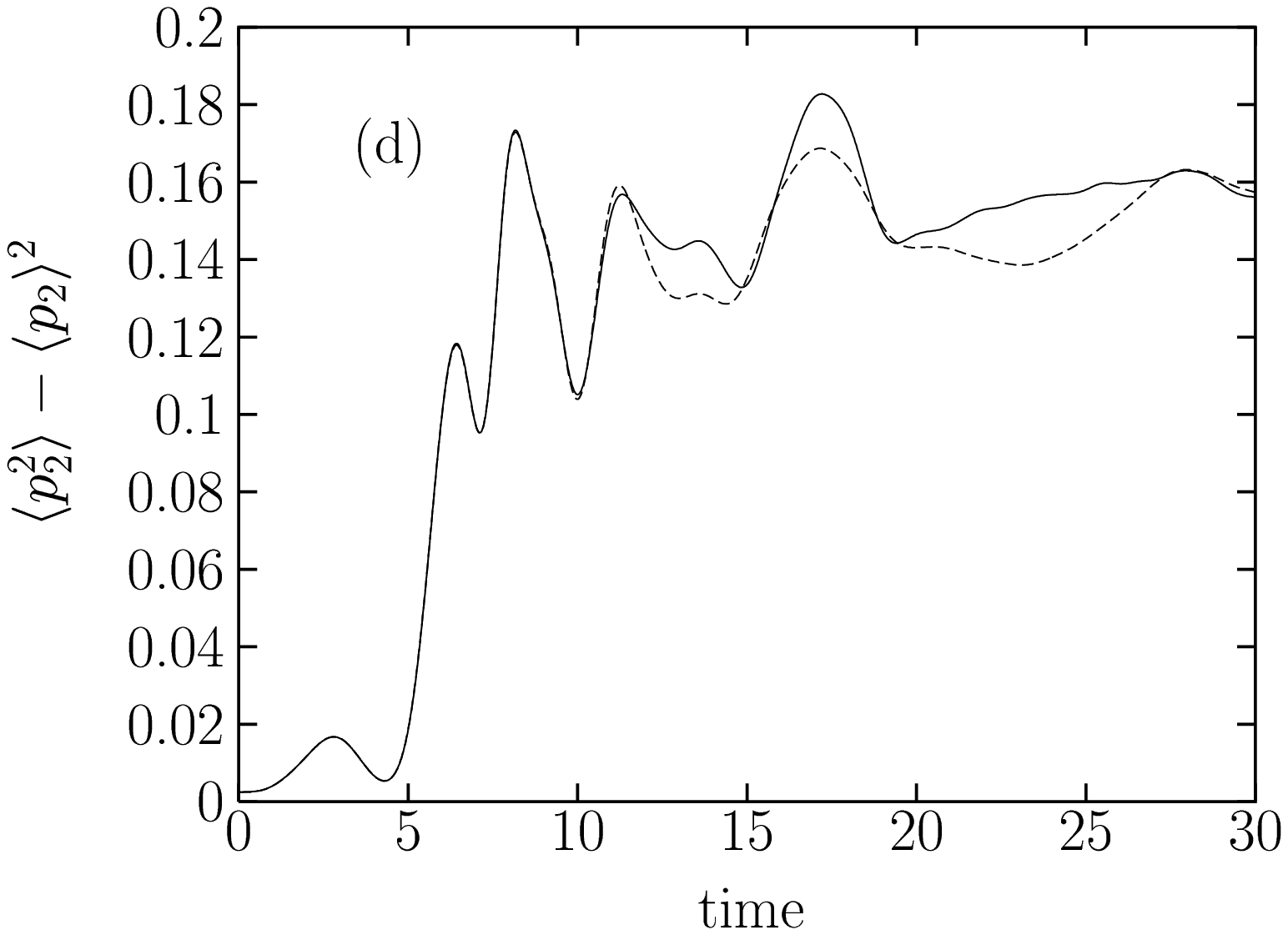,width=8.0cm}
\end{center}
\caption{Quantum classical comparison of the time dependence of
four second-order  statistical moments, i.e.,
$\langle q_{1}^{2}\rangle-\langle q_{1}\rangle^{2}$ in (a),
$\langle q_{2}^{2}\rangle-\langle q_{2}\rangle^{2}$ in (b),
$\langle p_{1}^{2}\rangle-\langle p_{1}\rangle^{2}$ in (c),
and $\langle p_{2}^{2}\rangle-\langle p_{2}\rangle^{2}$ in (d),
for times much larger than
the logarithmic break time $t_{b}$.
The initial distribution function  corresponds to that
in Figs. \ref{Fig2-3}a and \ref{Fig2-4}b, and $\hbar=0.005$.
Dashed lines denote classical results, solid lines are quantum
results.
The QCC shown here  during the complex relaxation process  $(0<t<12.0)$ is remarkable.
All variables are in dimensionless units.}
\label{Fig2-5}
\end{figure}

As already implied by a comparison between Eqs. (\ref{chi2q}) and
(\ref{foft}), what is directly responsible for the logarithmic break time
is not quantum classical difference in expectation values, but simply the
exponential increase of $f(t)$ which reflects the richness of
fragmentation of classical distribution functions. As such, it is
interesting to examine QCC in terms of some  observables. Figure
\ref{Fig2-5} displays the time dependence of four variances, i.e.,
$\langle q_{1}^{2}\rangle-\langle q_{1}\rangle^{2}$, $\langle
q_{2}^{2}\rangle-\langle q_{2}\rangle^{2}$, $\langle
p_{1}^{2}\rangle-\langle p_{1}\rangle^{2}$, and $\langle
p_{2}^{2}\rangle-\langle p_{2}\rangle^{2}$,  for both classical and
quantum dynamics.  The initial  state  corresponds to that used in Fig.
\ref{Fig2-3}a and Fig. \ref{Fig2-4}b and $\hbar$ still equals $0.005$,
allowing direct comparison with the previous results.

Interestingly, results in Fig. \ref{Fig2-5} show that QCC in these second
order moments is excellent for times up to  $t\approx 12.0$,  which is
much longer than the break time $t_{b} \sim 3.0$ identified in Fig.
\ref{Fig2-3}.  In particular, for times less than $t=2.5$,  both quantal
and  classical variances are seen to increase very rapidly, in exactly the
same manner. This rapid increase is  exponential in nature because, in
this regime, the sum of these variances (divided by $\hbar^{2}/2$) give
$\chi_{2c}^{2} $ (or $\chi_{2q}^{2} $), which indeed increases
exponentially on the average (see Fig. \ref{Fig2-3}a). This confirms a
published suggestion that the initial rapid increase of both quantal and
classical variances could be used to identify classical chaos from quantum
dynamics \cite{fox}. 
For times $2.5<t<12.0$
during which excellent QCC persists, the second order moments display a
complex evolution. For example, there is clearly a diffusive regime
between $t=5.0$ and $t=7.0$ with very large diffusion constants.  After
that, several very strong oscillations (particularly in Figs.
\ref{Fig2-5}b-d) can be observed, suggesting that both the quantal and
classical distribution functions alternate between a high degree of
delocalization and  a certain degree of localization over the entire
accessible phase space (this oscillatory behavior also suggests that the
QCC time scale here should scale as a power of $\hbar$, rather than scale
as $\ln\hbar$). The impressive QCC in this complex relaxation pattern is
in sharp contrast to the results in Figs. \ref{Fig2-3}a and \ref{Fig2-4}b,
where quantum classical differences, as quantitatively described by
$[\chi_{2c}^{2}(t)-\chi_{2q}^{2}(t)]$ and predicted by $f(t)$, are already
very large for $3.5<t<4.0$. For even later times ($t>12.0$),  Fig.
\ref{Fig2-5} shows that  quantum classical discrepancies begin to show up,
but still with very  similar trends in the oscillations.

These results indicate that the logarithmic break time may be quite
irrelevant to QCC when viewed from the perspective of some low order
statistical moments\cite{chris2}.  This is the case since considering low order
statistical moments is equivalent to projecting the quantal and classical
distribution functions onto  a much smaller subspace.  In this projection,
or coarse-graining procedure, all the information encoded in an infinite
number of higher order statistical moments is lost.  An accurate QCC is
thus restored due to the loss of detailed information.

The results here also give further support to  the idea of applying
classical propagation methods to quantum distribution functions, as a
means of approximating  the true quantum mechanics \cite{jaffe21}. That
is, chaos may not cause a rapid failure of classical dynamics simulations
if one is only interested in expectation values of some observables. More
importantly, the results provide more insights into a fundamental issue
regarding the role of decoherence in QCC \cite{zurek2,casati}. On the one
hand, our results here show that during a relaxation process whose time
scale is much larger than the logarithmic break time, decoherence may
still be unnecessary in order to ensure QCC in  low order statistical
moments. Thus, at least in our model system, decoherence effects on
correspondence in expectation values of low order statistical moments, if
any,  should be considered {\em after} the relaxation process is
essentially completed, e.g., after $t=12.0$ in Fig. \ref{Fig2-5}. This
point supports the argument of Casati {\it et al.} \cite{casati},
extending their considerations from one-dimensional kicked systems to
conservative systems, and agrees with the recent work \cite{ballentine01}
by Emerson and Ballentine. On the other hand, as shown in our previous
work using the same model system \cite{gong},  QCC for much larger time
scales can be much improved with  the introduction of decoherence.

\section{Summary}
\label{s2-6}

We have studied the issue of QCC in conservative chaotic systems in detail
using a phase space distribution function  approach.  The nature of QCC in
three different regimes is exposed.  In particular:  (i) the short time
increase rate of phase space structure is studied in connection with
finite-time Lyapunov exponents; (ii) a simple and enlightening description
of the break regime of QCC is obtained, by successfully accounting for the
classical phase space structure which is beyond the quantum description;
and (iii) excellent QCC in low order statistical moments is shown to
persist during a complex relaxation process, whose time scale is much
larger than the logarithmic break time.

\vspace{0.5cm} {\bf Acknowledgements}: This work was supported by the
Natural Sciences and Engineering Research Council of Canada.   We thank
Prof. Arjendu Pattanayak for useful discussions.

\appendix
\section{Derivation of Eq. (36)}

In this appendix we give a rather detailed derivation of Eq. (\ref{chi2expansion1})
using Eqs. (\ref{chi2t}), (\ref{Mtaylor}), (\ref{Md}) and (\ref{distri}).
We begin with the derivation  by substituting  Eq. (\ref{Mtaylor}) and
the initial distribution function $\rho_{0}[\qj(0), \qi(0)]$ [see Eq. (\ref{distri})]
into Eq. (\ref{chi2t}), evaluating the first derivatives of $\rho_{0}[\qi(0),\qj(0)]$,
and finally rescaling all integration
variables by a factor of $1/\sqrt{2}$. One then obtains
\begin{eqnarray}
\chi_{2c}^{2}(t)& = & \frac{1}{2}
\int d\qj(0)\rho_{0}[\qj(0), \qi(0)] \nonumber \\
&& \times \sum_{j}\left| \sum_{k} \alpha_{k}\left[A_{jk} +
\sum_{m}\frac{1}{\sqrt{2}}B_{jkm}\delta\qj_{m}(0)\right.\right.  \nonumber \\
&&  +
\sum_{mn}\left.\left.\frac{1}{4} C_{jkmn}\delta\qj_{m}(0)\delta\qj_{n}(0)+ O\left((\delta\qj(0))^{3}\right)\right]\right|^{2},
\label{chic1}
\end{eqnarray}
where \begin{eqnarray}
\alpha_{1}& = & \left[-\frac{p_{1}(0)-\overline{p}_{1}(0)}{\sigma_{p_{1}}^{2}}\right], \
\alpha_{2} = \left[-\frac{p_{2}(0)-\overline{p}_{2}(0)}{\sigma_{p_{2}}^{2}}\right], \nonumber
\\
 \alpha_{3}& = & \left[\frac{q_{1}(0)-\overline{q}_{1}(0)}{\sigma_{q_{1}}^{2}}\right], \
 \alpha_{4} =  \left[\frac{q_{2}(0)-\overline{q}_{2}(0)}{\sigma_{q_{2}}^{2}}\right].
\end{eqnarray}
Since $A_{jk}$, $B_{jkm}$, and $C_{jkmn}$ do not depend upon $\qj(0)$,
and only the even functions of $\delta\qj(0)$
will contribute when integrated over $\rho_{0}[\qj(0), \qi(0)]$,
Eq. (\ref{chic1}) can be further reduced to
\begin{eqnarray}
\chi_{2c}^{2}(t) & = & \sum_{jkk'mm'}\left(\frac{1}{4}A_{jk}C_{jk'mm'}+\frac{1}{4}B_{jkm}B_{jk'm'}\right)
\langle \alpha_{k}\alpha_{k'}\delta\qj_{m}(0)\delta\qj_{m'}(0)\rangle_{0}
\nonumber \\
&& + \frac{1}{2}\int d\qj \rho_{0}[\qj(0), \qi(0])
\left(\sum_{j}|\sum_{k}A_{jk}\alpha_{k}|^{2}\right)
+ O\left({\bf \alpha}^{2}(\delta\qj(0))^{4}\right),
\label{ckexpression}
\end{eqnarray}
where $\langle\cdot\rangle_{0}$ means the ensemble average over the initial Gaussian
distribution.

To further simplify the expression for $\chi_{2c}^{2}(t)$,
we make a change of the integration variables in the second term on the right side of
Eq. (\ref{ckexpression}), i.e.,
\begin{eqnarray}
-(p_{1}-\overline{p}_{1})&\rightarrow &\frac{\sigma_{p_{1}}}{\sigma_{q_{1}}} (q-\overline{q}_{1}), \
-(p_{2}-\overline{p}_{2})\rightarrow \frac{\sigma_{p_{2}}}{\sigma_{q_{2}}}(q_{2}-\overline{q}_{2}), \nonumber
 \\
(q_{1}-\overline{q}_{1})&\rightarrow & \frac{\sigma_{q_{1}}}{\sigma_{p_{1}}}(p_{1}-\overline{p}_{1}), \
(q_{2}-\overline{q}_{2})\rightarrow \frac{\sigma_{q_{2}}}{\sigma_{p_{2}}}(p_{2}-\overline{p}_{2}).
\end{eqnarray}
Note that the Jacobi matrix of this transformation is a unity matrix and it has no effect
on the form of  $\rho_{0}[\qj(0), \qi(0)]$.  As a result of this coordinate
transformation we have
$\alpha_{k}\rightarrow \delta\qj_{k}(0)/(\sigma_{q_{1}}\sigma_{p_{1}})$
=$\delta\qj_{k}(0)/(\sigma_{q_{2}}\sigma_{p_{2}})$.
This makes it possible to reexpress the second term on the right side of Eq. (\ref{ckexpression})
by use of Eq. (\ref{Md}).
Specifically,
\begin{eqnarray}
& &  \frac{1}{2}\int d\qj(0) \rho_{0}[\qj(0), \qi(0)]
\left(\sum_{j}|\sum_{k}A_{jk}\alpha_{k}|^{2}\right)  \nonumber \\
&=& \frac{1}{2(\sigma_{q_{1} }\sigma_{p_{1}})^{2}}
\int d\qj(0) \rho_{0}[\qj(0),
\qi(0)]
\left(\sum_{j}|\sum_{k}A_{jk}\delta\qj_{k}(0)|^{2}\right) \nonumber \\
&=& \frac{1}{2(\sigma_{q_{1}}\sigma_{p_{1}})^{2}}\left[
\sum_{j}\left(\langle(\qj_{j}(t))^{2}\rangle_{0}-\langle \qj_{j}(t)\rangle_{0}^{2}\right)
 -\sum_{jkl}B_{jkl}\langle\delta\qj_{j}(t)\delta\qj_{k}(0)
 \delta\qj_{l}(0)\rangle_{0}\right.
\nonumber \\
&& +\frac{1}{4}\sum_{jklk'l'}B_{jkl}B_{jk'l'}\langle \delta\qj_{k}(0)
\delta\qj_{k'}(0)
\delta\qj_{l}(0)\delta\qj_{l'}(0)\rangle_{0} \nonumber \\
& & -\left.\frac{1}{3} \sum_{jklm}C_{jklm}\langle\delta\qj_{j}(t)
\delta\qj_{k}(0)\delta\qj_{l}(0)\delta\qj_{m}(0)\rangle_{0}
 + O\left((\delta\qj(0))^{6}\right)\right].
\end{eqnarray}
The second  and the fourth terms in the above expression are a linear function of $\delta\qj(t)$
and they can be further transformed
into some functions of $\delta\qj(0)$
by using Eq. (\ref{Md}) a second time.
We then get
\begin{eqnarray}
&& \frac{1}{2}\int d\qj(0) \rho_{0}[\qj(0), \qi(0)]
\left(\sum_{j}|\sum_{k}A_{jk}\alpha_{k}|^{2}\right) \nonumber \\
& =&  \frac{1}{2(\sigma_{q_{1}}\sigma_{p_{1}})^{2}}
\sum_{j}\left[\langle(\qj_{j}(t))^{2}\rangle_{0}-\langle\qj_{j}(t)\rangle_{0}^{2}\right]
\nonumber \\
& &- \frac{1}{2(\sigma_{q_{1}}\sigma_{p_{1}})^{2}}
\sum_{jkk'll'}(\frac{1}{4}B_{jkl}B_{jk'l'}+\frac{1}{3}A_{jk}C_{jk'll'})\langle
\delta\qj_{k}(0)\delta\qj_{k'}(0)\delta\qj_{l}(0)\delta\qj_{l'}(0)
\rangle_{0} \nonumber \\
& &+  \ \frac{1}{(\sigma_{q_{1}}\sigma_{p_{1}})^{2}} O\left((\delta\qj(0))^{6}\right).
\label{secondterm}
\end{eqnarray}
Finally,  inserting  Eq.
(\ref{secondterm}) into Eq. (\ref{ckexpression}) yields
Eq. (\ref{chi2expansion1}).

\end{document}